\documentclass[
 reprint,
 amsmath,amssymb,
 aps,preprintnumbers
]{revtex4-1}

\usepackage{graphicx}
\usepackage{dcolumn}
\usepackage{bm}
\usepackage{amsmath,amssymb,graphicx,bbm,mathrsfs,nicefrac}
\usepackage{tikz}
\usepackage{amsmath}
\usepackage{amssymb}
\usepackage{graphicx,textcomp,float,gensymb,wrapfig, enumitem,comment,dsfont,framed,slashed,appendix,wrapfig,xcolor}
\usepackage{ bbold }
\usepackage{braket} 
\usepackage{ mathrsfs }
\usepackage{etoolbox,hyperref,makecell}
\usepackage{feynmp}
\usetikzlibrary{arrows.meta}
\usepackage{empheq}
\usepackage{enumitem}
\setenumerate[1]{label=(\arabic*)}
\newcommand{\bea}{\begin{eqnarray}}  
\newcommand{\eea}{\end{eqnarray}}

\newcommand{\nc}{\newcommand}

\nc{\beq}{\begin{equation}}
\nc{\eeq}{\end{equation}}
\nc{\barray}{\begin{eqnarray}}
\nc{\earray}{\end{eqnarray}}
\nc{\barrayn}{\begin{eqnarray*}}
\nc{\earrayn}{\end{eqnarray*}}
\nc{\bcenter}{\begin{center}}
\nc{\ecenter}{\end{center}}
\nc{\mc}{\mathcal}
\nc{\er}[1]{(\ref{eq:#1})}
\nc{\onehalf}{\frac{1}{2}} 
\nc{\partialbar}{\bar{\partial}}
\nc{\psit}{\widetilde{\psi}}
\nc{\hc}{\mbox{H.c.}}
\nc{\ev}{\;\mathrm{eV}}
\nc{\mev}{\;\mathrm{MeV}}
\nc{\gev}{\;\mathrm{GeV}}
\nc{\kev}{\;\mathrm{keV}}
\nc{\tev}{\;\mathrm{TeV}}

\def\chii0{\chi_i^0}
\def\chij0{\chi_j^0}

\newcommand{\gsim}{\lower.7ex\hbox{$\;\stackrel{\textstyle>}{\sim}\;$}}
\newcommand{\lsim}{\lower.7ex\hbox{$\;\stackrel{\textstyle<}{\sim}\;$}}
\nc{\ttbar}{t\bar t}

\newcommand{\fref}[1]{Fig.~\ref{#1}}
\newcommand{\eref}[1]{Eq.~(\ref{#1})}

\newcommand{\aref}[1]{Appendix~\ref{#1}}

\newcommand{\cref}[1]{Chapter~\ref{#1}}

\usepackage[utf8]{inputenc}

\begin{document}

\title{
Resurrecting the Fraternal Twin WIMP Miracle 
}

\author{David Curtin}
\email{dcurtin@physics.utoronto.ca}
\affiliation{Department of Physics, University of Toronto, Canada}
 
 \author{Shayne Gryba}
\email{sgryba@physics.utoronto.ca}
\affiliation{Department of Physics, University of Toronto, Canada}

 \author{Dan Hooper}
\email{dhooper@fnal.gov}
\affiliation{Fermi National Accelerator Laboratory, Theoretical Astrophysics Group, Batavia, IL 60510, USA}
\affiliation{University of Chicago, Kavli Institute for Cosmological Physics, Chicago, IL 60637, USA}
\affiliation{University of Chicago, Department of Astronomy and Astrophysics, Chicago, IL 60637, USA}

\author{Jakub Scholtz}%
\email{jakub.scholtz@unito.it}
\affiliation{Universit\`a di Torino, via P. Giuria 1, 10125, Torino, Italy}
\affiliation{Institute for Particle Physics Phenomenology, Durham University, Durham, DH1 3LE, United Kingdom}

\author{Jack Setford}%
\email{jsetford@physics.utoronto.ca}
\affiliation{Department of Physics, University of Toronto, Canada}

\date{\today}

\preprint{FERMILAB-PUB-21-294-T}

\begin{abstract}

In Twin Higgs models which contain the minimal particle content required to address the little hierarchy problem ({\it i.e.}~fraternal models), the twin tau has been identified as a promising candidate for dark matter. In this class of scenarios, however, the elastic scattering cross section of the twin tau with nuclei exceeds the bounds from XENON1T and other recent direct detection experiments. 
In this paper, we propose a modification to the Fraternal Twin Higgs scenario that we call $\mathbb{Z}_2$FTH, incorporating visible and twin hypercharged scalars (with $Y = 2$) which break twin electromagnetism.
This leads to new mass terms for the twin tau that are unrelated to its Yukawa coupling, as well as additional annihilation channels via the massive twin photon. 
We show that these features make it possible for the right-handed twin tau to freeze out with an acceptable thermal relic abundance while scattering with nuclei at a rate that is well below existing constraints. Nonetheless, large portions of the currently viable parameter space in this model are within the reach of planned direct detection experiments. 
The prospects for indirect detection using gamma rays and cosmic-ray antiprotons are also promising in this model. Furthermore, if the twin neutrino is light, the predicted deviation of $\Delta N_\mathrm{eff} \approx 0.1$ would be within reach of Stage 4 CMB experiments.
Finally, the high luminosity LHC should be able to probe the entire parameter space of the $\mathbb{Z}_2$FTH model through charged scalar searches. We also discuss how searches for long-lived particles are starting to constrain Fraternal Twin Higgs models.

\end{abstract}

\pacs{Valid PACS appear here}
\maketitle

\section{Introduction}

The hierarchy problem and the unknown nature of dark matter have long motivated searches for new physics at colliders and direct detection experiments. 
Proposed solutions to the hierarchy problem involve new states with masses near the weak scale which stabilize the mass of the Higgs boson. The problem of dark matter is also suggestive of new physics near the weak scale, motivated by the fact that a stable particle with a weak-scale mass and an annihilation cross section similar to that associated with the Standard Model (SM) weak interaction will freeze out of equilibrium in the early universe with a relic abundance similar to the measured density of dark matter. This so-called ``WIMP miracle'' has inspired an enormous range of dark matter models over the past several decades (for a historical review, see Ref.~\cite{Bertone:2016nfn}).
Naturally, any physics beyond the SM that is capable of solving both of these mysteries would be particularly well-motivated. This has long bolstered interest in supersymmetry~\cite{Martin:1997ns}, which requires each fermion to be accompanied by a bosonic partner (and vice versa) with identical gauge charges and couplings, thereby canceling quadratically divergent contributions to the Higgs mass. Some of these superpartner particles, in particular the neutralinos, are weakly interacting, and thus could serve as a dark matter candidate in the form of a weakly interacting massive particle (WIMP).
Naturalness considerations, however, lead one to expect the superpartners of the top quark to be light enough to be produced at a high rate at the Large Hadron Collider (LHC), in tension with the null results presented by the ATLAS and CMS collaborations~\cite{Aaboud:2017ayj,Aaboud:2017aeu,Aaboud:2018zjf,Sirunyan:2018lul,Sirunyan:2017leh,Sirunyan:2017kiw}. Similarly, across most of the supersymmetric parameter space, the lightest neutralino is predicted to scatter with nuclei at a rate that is excluded by direct detection experiments~\cite{Aprile:2018dbl,Akerib:2016vxi,Cui:2017nnn}. In light of these considerations, it is well motivated to consider alternatives to supersymmetry that can address the hierarchy problem and provide a dark matter candidate without conflicting with the constraints produced by the LHC and direct detection experiments.

Several scenarios have been proposed in which the quadratic contributions to the Higgs boson mass are cancelled by particles without SM gauge charges, constituting a paradigm known as ``Neutral Naturalness''~\cite{Chacko:2005pe,Barbieri:2005ri,Chacko:2005vw,Burdman:2006tz,Cai:2008au,Poland:2008ev,Craig:2015pha, Cohen:2018mgv, Cheng:2018gvu}. The most well-known and perhaps most promising of these models are those which fall within the Twin Higgs framework~\cite{Chacko:2005pe,Chacko:2005vw,Chacko:2005un,Barbieri:2005ri,Perelstein:2005ka,Craig:2013fga, Craig:2015pha}. In Twin Higgs models, a discrete $\mathbb{Z}_2$ symmetry stabilizes the Higgs mass at one-loop to solve the little hierarchy problem, with supersymmetry or compositeness solving the full hierarchy problem in the UV completion~\cite{Falkowski:2006qq,Chang:2006ra,Craig:2013fga,Craig:2016kue, Katz:2016wtw,Badziak:2017syq,Badziak:2017kjk,Badziak:2017wxn,Geller:2014kta,Barbieri:2015lqa,Low:2015nqa}.
This symmetry does not commute with SM gauge charges, giving rise to a twin sector of particles that have the same spin as their SM counterparts, but that are charged under twin versions of the SM forces. 
Since the twin tops are singlets under the SM strong force, this mechanism naturally evades standard LHC searches for top partners.

Just as supersymmetry must be broken in the infrared, the discrete symmetry of Twin Higgs models must be softly broken in order to make this framework a realistic theory of nature. 
In the perfect $\mathbb{Z}_2$-symmetric limit, the lightest Higgs mass eigenstate is an equal admixture of the visible and twin sector Higgs bosons, in significant conflict with measurements of the Higgs couplings~\cite{Burdman:2014zta, ATLAS:2020qdt, Sirunyan:2018koj}. Some source of $\mathbb{Z}_2$-breaking must therefore lift the twin Higgs vacuum expectation value (VEV), $f$, to a value that is a few times larger than that of the visible Higgs, typically $f/v \sim 3-5$. As a result, the 125 GeV Higgs boson contains a $\sim v^2/f^2 \sim \mathcal{O}(10\%)$ admixture of the twin sector Higgs, similar to the degree of tuning that the $\mathbb{Z}_2$ breaking must satisfy (though that tuning can be reduced in non-minimal constructions~\cite{Beauchesne:2015lva,Harnik:2016koz, Csaki:2019qgb}). 
This Higgs portal, which is essential for the solution to the little hierarchy problem, provides a means by which this class of models can be probed at the LHC and future colliders. In particular, the mixing predicts universal deviations among the Higgs couplings~\cite{Burdman:2014zta}, which could be detected at the high-luminosity LHC and would be exhaustively probed at future lepton colliders~\cite{Dawson:2013bba, Craig:2013xia}. 
The Higgs portal also allows for exotic Higgs decays into the twin sector, which may result in the production of long-lived particles~\cite{Craig:2015pha, Curtin:2015fna}.

In the unbroken $\mathbb{Z}_2$-limit, Twin Higgs models include a massless twin photon and three light twin neutrinos, all of which are problematic for cosmology. In particular, these particles are collectively predicted to contribute to the energy density of radiation at a level corresponding to $\Delta N_{\rm eff} \sim 5$, while CMB measurements have produced an upper limit of $\Delta N_{\rm eff} < 0.23 ~(2 \sigma)$~\cite{Aghanim:2018eyx}.
Realistic Twin Higgs models must therefore either eliminate these light degrees of freedom by introducing a hard breaking of the $\mathbb{Z}_{2}$ symmetry (while retaining the cancellation between the top and twin top contributions to the Higgs mass)~\cite{Craig:2015pha,Garcia:2015loa}, or by introducing an asymmetric reheating mechanism that increases the temperature of the visible sector relative to the twin sector, thereby lowering $\Delta N_{\rm eff}$~\cite{Chacko:2016hvu,Craig:2016lyx}.%

The Fraternal Twin Higgs (FTH) model~\cite{Craig:2015pha} is a particularly simple and appealing representation of the first possibility. The twin sector in the most minimal version of this model consists of $t', b', \tau'$, and $\nu_{\tau}'$, which are charged under twin QCD and twin weak interactions in analogy to the SM. Twin electromagnetism is either not gauged in the most minimal version of this model, or is broken at a high scale. For the most $\mathbb{Z}_2$-preserving version of this particle content, the masses of the twin fermions are set by SM-like Yukawa couplings to the twin Higgs, making them heavier than SM fermions by  a factor of $f/v$. Alternatively, one could break the $\mathbb{Z}_2$ symmetry further, allowing for us to treat $m_{b'}, m_{\tau'}$, and $m_{\nu_\tau'}$ as free parameters (although the twin top mass must remain fixed to $(f/v) \times m_t$ in order to stabilize the mass of the light Higgs).

By virtue of an accidental global $U(1)$ symmetry, the twin tau is stable in the FTH model, making it a potential candidate for dark matter~\cite{Garcia:2015loa,Craig:2015xla}. 
(See~\cite{Farina:2015uea,
Garcia:2015toa,
Farina:2016ndq,
Freytsis:2016dgf,
Cheng:2018vaj,
Hochberg:2018vdo,
Badziak:2019zys,
Koren:2019iuv,
Terning:2019hgj,
Beauchesne:2020mih,
Ahmed:2020hiw,
Curtin:2021alk, Ritter:2021hgu} for other dark matter candidates within the Twin Higgs framework.) 
In particular, the twin tau annihilates primarily to twin neutrinos or twin bottoms through the twin weak interaction (which is a copy of the SM weak interaction, with $m_{W'} = (f/v) \times m_{W}$), yielding an acceptable thermal relic density for masses in the range of $m_{\tau'} \sim 50 - 150 \gev$.
This ``Fraternal Twin WIMP Miracle" 
plays out almost entirely within the hidden sector: annihilations into the visible sector through the Higgs portal play a subdominant role, except near $m_{\tau'} \sim m_h/2$. However, the Higgs portal does lead to irreducible signatures. 
Since the entirety of the twin tau mass comes from its Yukawa coupling to the twin Higgs, the twin tau is predicted to have a sizable elastic scattering cross section with nuclei through the Higgs portal. 
Therefore, constraints from XENON1T and other recent direct detection experiments~\cite{Aprile:2018dbl,Akerib:2016vxi,Cui:2017nnn} have by now excluded the entire range of parameter space that yields an acceptable relic abundance in this model.

A very minimal modification of the FTH model is to reintroduce the twin photon, $\gamma'$. Recently, a new mechanism was proposed for the Mirror Twin Higgs model to spontaneously break the $\mathbb{Z}_2$ symmetry in the Higgs sector by extending it to include scalars charged under visible and twin hypercharge~\cite{Batell:2019ptb}. One of these new scalars, by definition in the twin sector, acquires a VEV, giving a mass to the twin photon and generating the soft $\mathbb{Z}_2$-breaking Higgs mass term that raises $f$ above $v$. Depending on the hypercharge of the scalar, various additional twin fermion mass terms can also be generated through this mechanism.

In this paper, we apply the mechanism proposed in Ref.~\cite{Batell:2019ptb}, implemented with $Y=2$ (twin) hypercharge scalars, to a maximally $\mathbb{Z}_2$-symmetric FTH model. We demonstrate that this ``$\mathbb{Z}_2$FTH" model with spontaneously broken twin hypercharge provides a viable dark matter candidate in the form of a dominantly right-handed twin tau with a mass on the order of $\sim \mathcal{O}(100 \gev)$.
Unlike dark matter in the conventional FTH model, our scenario is consistent with existing direct detection constraints. 
This is made possible due to the new $\tau'$ Majorana mass terms that are unrelated to the elastic scattering cross section, as well as new annihilation processes mediated by the twin photon. 
The $\mathbb{Z}_2$FTH scenario leads to a variety of potentially observable signals, including those at future direct detection experiments, exotic Higgs decay searches, CMB measurements, and indirect dark matter searches.

\section{A New Fraternal Twin WIMP Miracle}

In Ref.~\cite{Batell:2019ptb}, Batell and Verhaaren augmented the standard Twin Higgs model with new scalars, $\Phi = (\phi, \phi')$, in the visible and twin sectors, respectively. The spontaneous $\mathbb{Z}_2$ breaking can be understood from a simplified $\Phi$-only potential,
\begin{equation}
\label{eq:simplified_potential}
V_\Phi = -\mu_\Phi^2 |\Phi|^2 + \lambda_\Phi |\Phi|^4 + \delta_\Phi (|\phi|^4 + |\phi'|^4) \ ,
\end{equation}
which is $\mathbb{Z}_2$ symmetric and obeys a $U(2)$ symmetry in the $\delta_\Phi \to 0$ limit. 
If $\delta_\Phi < 0$, one of the scalars has no VEV, while the other scalar (by definition $\phi'$) acquires a VEV, $f_\phi$.
The resulting physical degrees of freedom are as follows:
\begin{itemize}
\item The twin hypercharge gauge boson, $B'$, acquires a mass of $m_{B'} = \sqrt{2} Y g' f_\phi$, where $Y$ is the hypercharge carried by the new scalars and $g'$ is the twin hypercharge gauge coupling.

\item The radial component of $\phi'$, labeled $\rho$, is the radial mode of the approximate $U(2)$ breaking with a mass of $m_{\rho}\sim \sqrt{\lambda_\Phi} f_\phi$. The angular mode is eaten by the $B'$.

\item The visible sector charged scalar, $\phi$, is a pseudo-Goldstone boson with a mass given by $m_{\phi} \sim \sqrt{\delta_\Phi} f_\phi$. Note that $\phi$ can be much lighter than the twin sector radial mode, $\rho$.
\end{itemize}
This potential can be integrated into the Twin Higgs scalar sector. The $\mathbb{Z}_2$-symmetric coupling between the hypercharge breaking scalars and the visible and twin electroweak doublet Higgs bosons, 
\begin{equation}
\delta_{H \Phi} (|H|^2 - |H'|^2)(|\phi|^2 - |\phi'|^2) \ ,
\end{equation}
then generates the $\mathbb{Z}_2$-breaking Higgs mass term once $\phi'$ acquires its VEV. 
The mass spectrum of the twin electroweak gauge bosons is straightforward, with the caveat that we always define $\gamma'$ as the admixture of $B'$ and $W'^3$ that is predominantly $B'$, while $Z'$ is always predominantly $W'^3$. Thus $m_{\gamma'} \approx m_{B'}$ when $m_{B'} \gg m_{Z'}$ or $m_{B'} \ll m_{Z'}$, with a discontinuous level crossing in the spectrum around $f_{\phi} \sim f$, when the twin photon and $Z'$ are similar in mass.

This mechanism accomplishes the required breaking of the $\mathbb{Z}_2$ symmetry in the scalar sector of the Twin Higgs model in a simple and elegant way. Note that the new scalars are protected from quadratic divergences via the same twin mechanism as the SM Higgs.
It also makes interesting physical predictions, including the existence of the electrically charged scalar in the visible sector, $\phi$. If this scalar were stable, it would be an unacceptable charged relic. To evade this problem, we require that $Y = 1$ or $Y = 2$, allowing for Yukawa couplings that enable $\phi$ to decay to fermions in each sector. 
In the $Y=1$ case, breaking of twin hypercharge would allow twin-electric charge to be violated by units of $\Delta Q' = 1$, making the twin tau unstable. Since we want to identify $\tau'$ as a dark matter candidate, we focus on the case of $Y=2$, where the lightest $\tau'$ mass eigenstate can be stable. 
The visible-sector $\phi$ must then be heavy enough to evade current bounds on new doubly-charged particles ($m_{\phi} \gtrsim 500 \, \gev$ \cite{CMS:2017pet}, assuming that $\phi$ decays exclusively to taus, or $\gtrsim 600-800 \gev$ if there are significant decays to electrons or muons). 
It is convenient to keep in mind that $m_{\gamma'} \approx f_\phi$ in this case (neglecting $Z'$-mixing).

With these considerations in mind, we define the \emph{maximally $\mathbb{Z}_2$-symmetric Fraternal Twin Higgs scenario} ($\mathbb{Z}_2$FTH) as follows. We begin with the FTH particle spectrum ({\it i.e.}\,\,only 3rd generation twin fermions), requiring all gauge and Yukawa couplings to take on their $\mathbb{Z}_2$ symmetric values ({\it i.e.}\,\,equal to the SM values, up to RGE effects). We then restore the twin photon, and add the $Y=2$ hypercharge breaking field to the scalar sector in order to generate the spontaneous $\mathbb{Z}_2$ breaking in the Higgs sector and give the twin photon a mass. 
If some dynamical mechanism were to remove the first two generations of twin fermions from the low-energy spectrum, this theory could be fully $\mathbb{Z}_2$ symmetric in the UV. Alternatively, the discrete symmetry might only apply to the gauge, scalar and third generation fermion sectors, as is the case in some proposed UV completions~\cite{Craig:2014roa}.
In any case, the absence of the 1st and 2nd generation twin fermions in our model has the fortunate side effect of solving the domain wall problem of spontaneous $\mathbb{Z}_2$ breaking that is found in the original implementation of this model~\cite{Batell:2019ptb}, since the loop-induced quartic couplings of the $\phi, \phi'$ scalars are no longer symmetric (see Appendix~\ref{app:domainwalls} for details).
Therefore, once the asymmetry in the fermion content between the two sectors is established by some other means, this mechanism ensures that the sector containing less matter also has broken hypercharge and a higher Higgs VEV, establishing its identity as the ``twin'' sector.

The $\mathbb{Z}_2$FTH scenario features new mass terms for the twin tau, generated from interactions with $\phi^{\prime}$. 
In particular, there is a new Yukawa coupling for the charged right-handed tau field, $\bar{E}_\tau^{\prime}$, in the twin sector:
\begin{equation}
\label{e.LRR}
- \mathcal{L} \supset 
\lambda_\tau \phi' \bar E'_\tau \bar E'_\tau
 + \ \mathrm{h.c.},
\end{equation}
as well as new non-renormalizable interactions involving the charged left-handed lepton field, $L_\tau^{\prime}$, and Higgs doublet, $H^{\prime}$:
\begin{eqnarray}
\label{e.newyukawa}
\nonumber
 - \mathcal{L} 
\supset  
\frac{\xi_\tau}{\Lambda^2} \phi'^* (L_\tau' H'^\dagger)^2 + 
 \label{e.LLLLR}
\frac{\tilde c_\tau}{\Lambda^2} |\phi'|^2 L_\tau' {H'}^\dagger \bar E_\tau' 
 \ +  \mathrm{h.c.}, \\
 \end{eqnarray}
where $\Lambda$ is the scale at which the Twin Higgs scenario is UV completed. Note that analogous terms appear in the visible sector, but these do not lead to new mass terms because $\braket{\phi} = 0$. 
Once $H, H'$, and $\phi'$ acquire their VEVs, these terms give rise to a non-trivial twin tau mass matrix:
\begin{equation}
\mathcal{L} \supset -\frac{1}{2} \bar \psi_i (M_{\tau'})_{ij} \psi_j + \mathrm{h.c.} \ , \ \ 
M_{\tau'} = 
\left(
\begin{array}{cc}
m_{LL} & m_{LR}\\
m_{LR}  & m_{RR}
\end{array}
\right), \\ 
\label{e.Mtau}
\end{equation}
where
\begin{eqnarray}
\label{e.mLR}
m_{LR} &=& \frac{y_\tau}{\sqrt{2}} f + \frac{\tilde c_\tau}{\sqrt{2} }\frac{f_\phi^2}{\Lambda^2} f,
\\
\label{e.mLL}
m_{LL} &=& \xi_\tau \frac{f^2}{\Lambda^2} f_\phi,
\\
\label{e.mRR}
m_{RR} &=& 2 \lambda_\tau f_\phi, 
\end{eqnarray}
and $y_\tau \approx 0.01$ is the $\mathbb{Z}_2$-symmetric twin tau Yukawa coupling.
In general, the $\tau'$ mass eigenstates will be two Majorana fermions, $\tau'_1, \tau'_2$, with masses $m_{\tau'_1} < m_{\tau'_2}$.

Two effects are responsible for the stability of $\tau'_1$ and hence its suitability as a dark matter candidate. First, the high mass of the twin top can forbid hadronic $\tau'_1$ decays into twin sector states, even though $\tau'$ lepton number is no longer conserved due to twin hypercharge breaking. Second, the unbroken $U(1)_{\text{EM}}$ in the visible sector ensures the twin tau cannot decay into SM particles. 

We can now understand why the twin tau is a viable WIMP in our new $\mathbb{Z}_2$FTH scenario.
First, the twin photon enhances the twin tau annihilation cross section to both $\bar{b}' b'$ and $\bar{\nu}' \nu'$ final states, with the latter interaction arising because the $\phi'^2 B'_\mu B'^\mu$ coupling causes the mixing angle between the $B'_\mu$ and $W'^3_\mu$ to differ from the Weinberg angle.  We assume that $m_{b'} = (f/v)~ m_b$ to maximally respect the $\mathbb{Z}_2$ symmetry, and that the twin bottoms annihilate or decay to twin glueballs which subsequently decay to the SM through the Higgs portal~\footnote{Even though we assume twin bottom and glueball masses expected from the $\mathbb{Z}_2$ symmetry, their exact masses are not  important for the viability of this scenario, as long as they can annihilate or decay into the visible sector.}.
Secondly, for $m_{\gamma'} \gtrsim \mathcal{O}(100 \gev)$, we find that twin taus with $m_{\tau'_1} > m_{b'} \gtrsim15 \gev$ can freeze out with the desired relic density without being excluded by direct detection constraints. 
In order to raise $m_{\tau'}$ above the twin bottom mass, we make use of the new mass contributions shown in Eqns.~(\ref{e.mLR})-(\ref{e.mRR}). In contrast to the original FTH scenario, which relied on adjusted twin Yukawa couplings, these new mass terms make it possible to realize $\tau'_1$ as a dark matter candidate while fully respecting the $\mathbb{Z}_2$ symmetry. 

\begin{figure}[t]
\includegraphics[width=0.40 \textwidth]{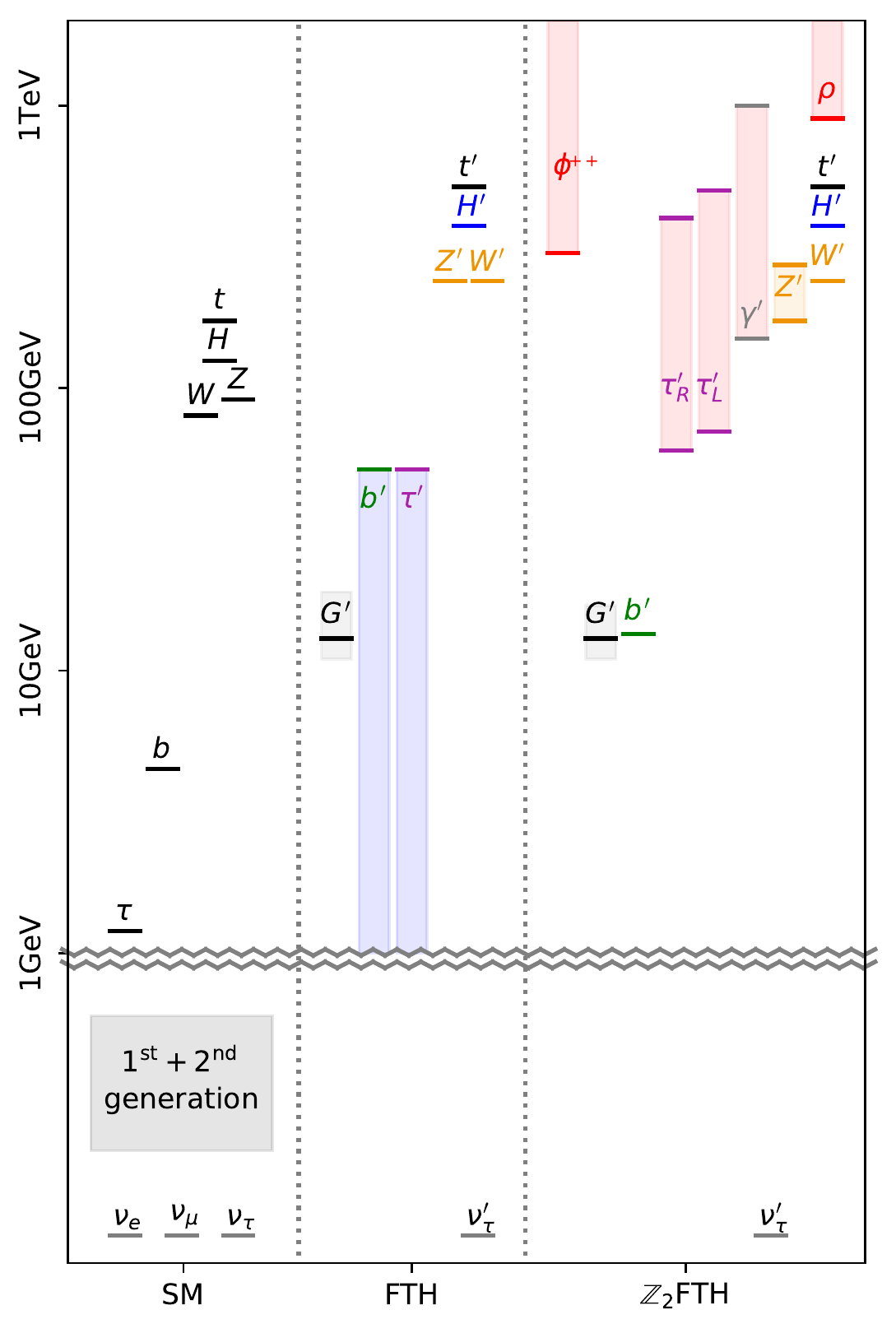}
\caption{The particle spectrum of the Standard Model (left), compared to that predicted in the original Fraternal Twin Higgs (FTH) model (center), and in the $\mathbb{Z}_2$-symmetric FTH model with  $\tau'_R$ dark matter  as presented in this paper (right). In original FTH model, the masses of twin bottom and tau are controlled by their $\mathbb{Z}_2$-breaking Yukawa couplings, with their allowed range represented in blue. In the $\mathbb{Z}_2$FTH, the masses of twin taus and twin photon are determined by their coupling to $\tilde{\phi}$, with their allowed ranges shown in red. Note that $\phi^{++}$ is in the visible sector.
}
\label{fig:spectrum}
\end{figure}

In principle, the lightest twin tau in the $\mathbb{Z}_2$FTH model could be mostly $\tau'_R$ (if $m_{LL} > m_{RR} \gg m_{LR}$) or mostly $\tau'_L$ (if $m_{RR} > m_{LL} \gg m_{LR}$). Alternatively, in the special case in which $\xi_\tau/\Lambda^2$ and $\lambda_\tau$ are small, the mass splitting between the two twin tau states could be very small, allowing us to treat the two twin tau states as a single pseudo-Dirac fermion. 
However, the $\tau'_L$ or pseudo-Dirac dark matter candidates have an effective Yukawa coupling to the 125 GeV Higgs that scales with their mass, $y_{\tau_1}^\mathrm{eff} \sim (m_{\tau_1}/f) \sin \vartheta $  (where $\sin \vartheta \approx v/f$). This is similar to the original Fraternal Wimp Miracle~\cite{Garcia:2015loa,Craig:2015xla} and is already excluded by direct detection.
On the other hand, $m_{RR}$ in Eqn. (\ref{e.mRR}) is entirely generated by the hypercharge breaking VEV. This allows a dominantly $\tau'_R$ dark matter candidate to be sufficiently heavy while also avoiding sizeable Higgs portal couplings.
We therefore focus on this case for the remainder of our study.

In Fig.~\ref{fig:spectrum}, we show a sketch of the particle spectrum expected in this new $\mathbb{Z}_2$FTH model, for the case in which the lightest twin tau is mostly $\tau'_R$, and compare this to the original FTH scenario.
To realize the $\tau'_R$ dark matter candidate, the $\xi_\tau$ interaction in \eref{e.mLL} must be sufficiently large to ensure $m_{LL} > m_{RR}$. This will generate a mass correction to $H'$ at loop order, which can spoil the Twin Higgs solution of the little hierarchy problem~\cite{Batell:2019ptb}. An extremely  generous upper bound for this coupling can be derived by requiring that this correction be no larger than gauge or top loop corrections, which implies $\xi_{\tau} \lsim \Lambda^2/(f f_\phi)$, but any sensible theory should have $m_{LL}$ significantly below this bound. 
Therefore our model must satisfy 
\begin{equation}
\label{e.mLLmax}
m_{\tau_1'} < m_{\tau_2'} < f.
\end{equation}
However, an even stronger constraint is obtained by requiring that the twin-hadronic decay $\tau'_1 \to \nu' b' t'$ is kinematically forbidden,
\begin{equation}
\label{e.mtau1max}
m_{\tau_1'} < m_{t'} + m_{b'}.
\end{equation}
We find that viable scenarios with the correct dark matter relic abundance can easily satisfy these requirements, as long as the scale of the UV completion, $\Lambda$, is around a few TeV.

The extended scalar sector  plays an important role in the dark matter phenomenology within this model.
The radial mode of the twin hypercharge breaking scalar, $\rho$, mixes slightly with the visible sector Higgs after $\mathbb{Z}_2$ breaking (see Ref.~\cite{Batell:2019ptb} for details). 
The  $h-\rho$ mixing angle is approximately given by 
\begin{equation}
\label{eq:sinalpha}
\sin \alpha \approx \frac{- f^2 m_h^2}{\sqrt{2} f_\phi v m_\rho^2}  \approx - 0.01 \left(\frac{f}{v}\right)^2 \left(\frac{\mathrm{TeV}}{m_\rho}\right)^{2} \left(\frac{300 \ \mathrm{GeV}}{f_\phi}\right) \ ,
\end{equation}
which is too small to be relevant for LHC phenomenology or to have an impact on the thermal freeze-out of the twin taus. However, this mixing will have important consequences for direct detection, making it  necessary to understand the possible size of $f_\phi$ and $m_\rho$.

The mixing term and the Higgs mass are parametrically related, since both are generated by the nonzero value of $f_\phi$ (recall that the visible sector Higgs is a pseudo-Goldstone boson whose mass arises from $\mathbb{Z}_2$-violating effects). Fixing the ratio between the off-diagonal terms and the Higgs mass term bounds the larger mass eigenvalue from below. 
The mass of the radial mode must therefore satisfy
\begin{equation}
\label{eq:rho_mass}
m_\rho - \frac{m_h^2}{m_\rho} \gtrsim \frac{1}{\sqrt{2}} \frac{m_h f^2}{v f_\phi}.
\end{equation}
For $f/v = 3$ and $f_\phi = 300$ GeV, this corresponds to $m_{\rho }\gsim 700 \gev$.
In addition, there is a maximum value for the mass of the radial mode, 
\begin{equation}
\label{e.mrhomax}
m_\rho \lsim 2 \sqrt{\lambda_\Phi^{\rm max}} f_\phi,
\end{equation}
where $\lambda^{\rm max}_{\Phi} = (2/3) \times 4\pi$ is the upper bound from unitarity considerations~\cite{Goodsell:2018tti}. 
In addition to bounding the acceptable range for $m_{\rho}$, these conditions also require that 
\begin{equation}
\label{e.fphimin}
f_\phi \gtrsim (60 \gev) \times  \frac{f}{v}  \ ,
\end{equation}
which further restricts the parameter space of our theory.

\begin{figure}
\begin{center}
\includegraphics[width=0.48 \textwidth]{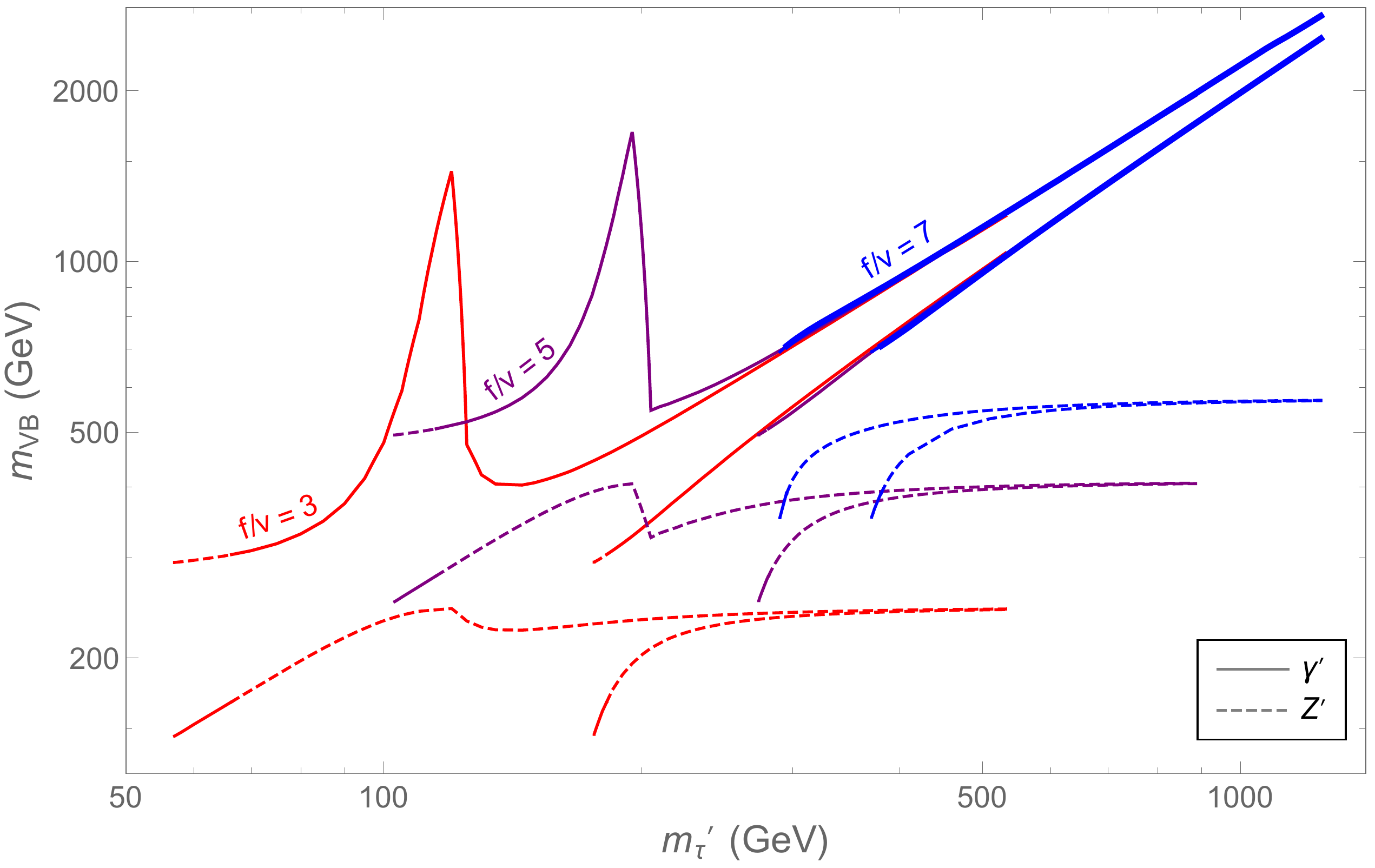}
\end{center}
\caption{
The values of $m_{\gamma'}$ and $m_{Z'}$ as a function of $m_{\tau'_1}$ that are required to obtain $\Omega_{\tau'_R} h^2 = 0.12$~\cite{Aghanim:2018eyx}, for the case in which the lightest twin tau is mostly $\tau_R'$. The gauge boson eigenstates are identified such that $\gamma'$ is always mostly $B'$ and $Z'$ is mostly $W'^3$. Results are shown for $f/v = 3, 5$ and 7. Note that the dark photon masses coincide for heavy $\tau'$. The maximum $\tau'_1$ mass is  given by \eref{e.mtau1max}.
}
\label{f.darkphotonmass}
\end{figure}

The relic abundance of the twin tau is determined in large part by its twin sector annihilation cross section through $Z'$ and $\gamma'$ exchange to twin bottoms and twin neutrinos (see \aref{a.xsecs}). In the special case of $m_{\tau'} \sim m_h/2$, annihilations through the Higgs portal into the visible sector can also play an important role. In Fig.~\ref{f.darkphotonmass}, we plot the values of the twin photon and twin tau masses that lead to the measured dark matter relic abundance, for the case in which the lightest twin tau is mostly $\tau_R'$ and for three representative values of $f/v$. Since the twin photon plays a crucial role in the determination of the relic abundance,  its minimum mass in turn dictates a minimum value of $m_{\tau_1'}$ for each case.

%%%%%%%%%%%%%%%%%%%%%%%%%%%%%%%%%%%%%%%%%%%%
%%%%%%%%%%%%%%%%%%%%%%%%%%%%%%%%%%%%%%%%%%%%
%%%%%%%%%%%%%%%%%%%%%%%%%%%%%%%%%%%%%%%%%%%%

\section{Experimental Signatures}

Although the values of $m_{\rho}$, $m_{LL}$ and $m_{LR}$ do not appreciably impact the relic abundance of our $\tau'_R$ dark matter candidate, they entirely determine the magnitude of its direct detection signal. 
The elastic scattering cross section with nuclei has two important contributions. 
First, the coupling of $\tau'_R$ with $\rho$, which in turn mixes with the Higgs, leads to an effective Yukawa coupling $\frac{1}{\sqrt{2}} y_{\tau_1}^\mathrm{eff,R} h \bar{\tau}'_1 \tau'_1$ that is given by 
\begin{equation}
y_{\tau'_1}^\mathrm{eff,R(\rho)} = \sqrt{2}\lambda_\tau \sin \alpha,
\end{equation}
with $\sin  \alpha$ defined in Eqn.~(\ref{eq:sinalpha})~\footnote{Note that the Majorana fermion nature of the $\tau'_1$ introduces an extra factor of 2 into the Feynman rule associated with this vertex.}. This scales as $\sim 1/m_\rho$ across the relevant range of $m_{\gamma'}$. Recall from the discussion in the previous section that the value of $m_{\rho}$ is bounded from both above and below, allowing us to bound this contribution to the dark matter's coupling to nuclei. 

A second contribution to the coupling is generated as a result of the mixing of $\tau'_R$ with $\tau'_L$, thereby accessing the usual Higgs portal. 
The contribution to this coupling is given by
\begin{align}
\label{e.yEffR}
y_{\tau'_1}^\mathrm{eff, R (\tau'_L)} 
  &= \frac{\sqrt{2}}{f} \frac{m_{LR}^2}{m_{LL}} \left(\frac{m_{RR}}{m_{LL}} \right)^2
  +  \mathcal{O} \left(\frac{m_{RR}}{m_{LL}}  \right)^3.
\end{align}
Since $m_{\tau'_2}$ is bounded by \eref{e.mLLmax}, this contribution to the direct detection signal can also be bounded from both above and below.

\begin{figure}
\begin{center}
\includegraphics[width=0.48 \textwidth]{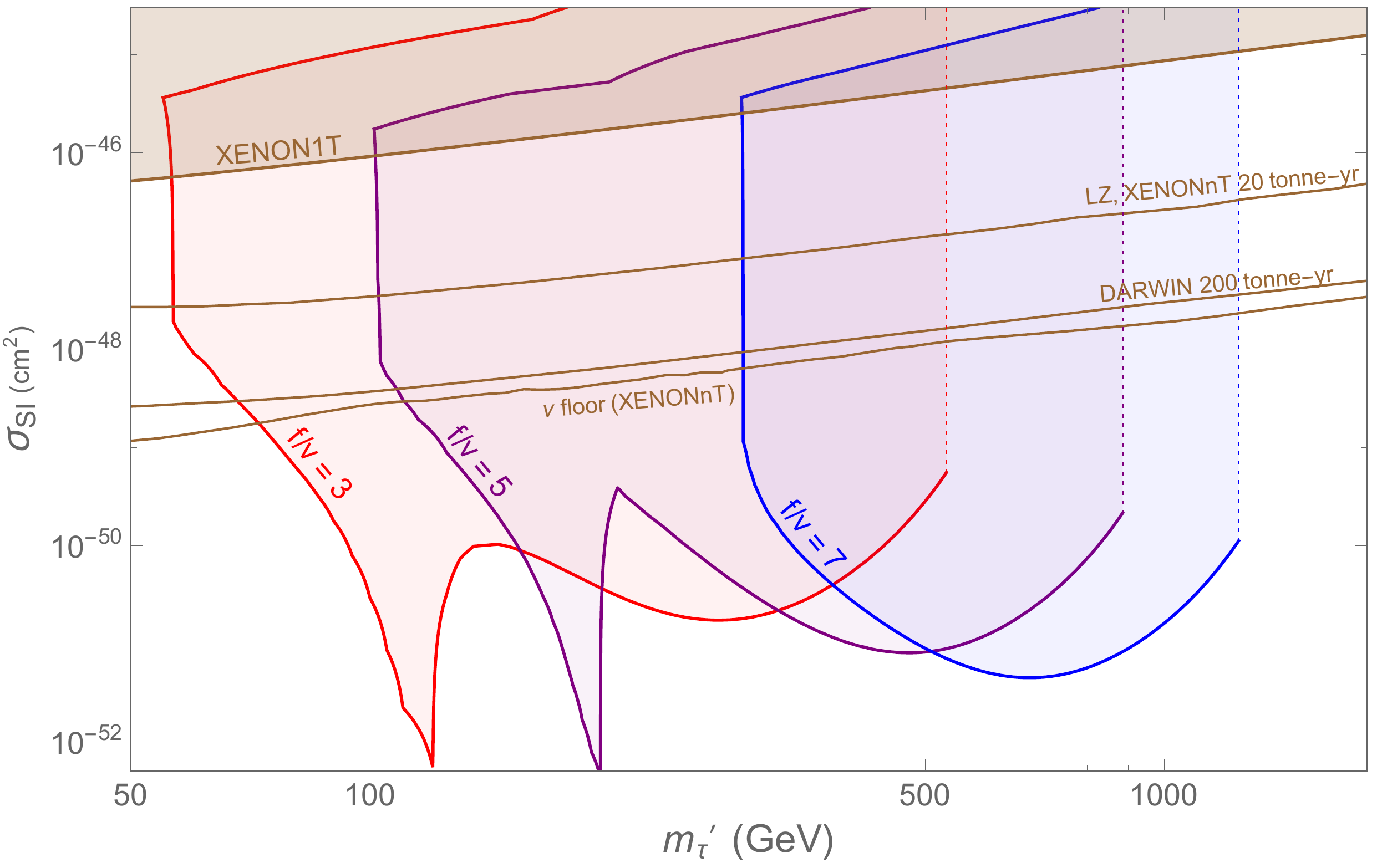}
\end{center}
\caption{The allowed range  in the $\mathbb{Z}_2$FTH model of the twin tau's spin-independent elastic scattering cross section with nuclei. The vertical dashed lines indicate the upper bound on $m_{\tau'_1}$ from requiring stability against twin hadronic decays.
Shown in brown are the current constraints from XENON1T~\cite{Aprile:2018dbl}, and the projected constraints from LUX-Zeplin (LZ)~\cite{Akerib:2015cja} and XENONnT \cite{Aprile:2015uzo} after 20 tonne-years of exposure, as well as from DARWIN~\cite{Aalbers:2016jon} with an exposure of 200 tonne-years. The neutrino floor curve was taken from Ref.~\cite{Aprile:2015uzo}.}
\label{f.directdetection}
\end{figure}

The range of the possible direct detection signal is depicted in the shaded regions of Fig.~\ref{f.directdetection}, for three different values of $f/v$. 
The upper boundary represents the largest possible scattering cross section, realized in the pseudo-Dirac limit, where $m_{LL} = m_{RR}$ and $m_{\tau'_2} - m_{\tau'_1} = 2 m_{LR}$, with $m_{LR}$ set to the smallest value possible for the minimal assumption of $\tilde c_{\tau} = 0$. Other values would not significantly affect the results. We also maximise $y_{\tau'_1}^\mathrm{eff, R (\rho)}$ by minimising the possible value of $m_\rho$ (see \eref{eq:rho_mass}), but its contribution is negligible compared to $y_{\tau'_1}^\mathrm{eff, R (\tau'_L)}$.

The lower boundary, representing to the smallest possible cross section, corresponds to the maximum value of $m_\rho$ (see \eref{e.mrhomax}), and to $m_{LL} \approx m_{\tau'_2}$ saturating the upper bound in \eref{e.mLLmax}.
The right boundary of the allowed $m_{\tau_1'}$ range is set by maintaining stability of $\tau'_1$ against twin-hadronic decays (see \eref{e.mtau1max}).
The left boundary is set by the minimum value of $m_{\gamma'}$ (see \eref{e.fphimin} and \fref{f.darkphotonmass}). 
For $\tau'_1$ masses near the minimum allowed value, the lower bound on the scattering cross section is entirely dominated by $\rho$ mixing, while for larger dark matter masses, mixing with $\tau_L'$ contributes significantly. 

From Fig.~\ref{f.directdetection}, it is clear that our twin tau dark matter candidate could potentially be discovered at upcoming underground dark matter experiments. In significant portions of the parameter space, however, the elastic scattering cross section is below the so-called ``neutrino floor'', making it possible that this candidate could escape direct detection. 
Fortunately, the prospects for indirect detection are also promising in this scenario. The twin tau annihilates in this model mostly into twin bottoms, at a rate given by the thermal annihilation cross section. These twin bottoms then form quirky bound states~\cite{Okun:1980kw, Okun:1980mu, Kang:2008ea,Craig:2015pha} which further annihilate or decay into twin glueballs that in turn decay into SM particles through the Higgs portal (see \aref{a.lhcllp}). The signals associated with this process could likely be used to place constraints on our scenario, and may be able to explain observed gamma-ray and cosmic-ray excesses~\cite{Daylan:2014rsa,Hooper:2010mq,Cuoco:2016eej,Cui:2016ppb,Cholis:2019ejx}. A quantitative assessment of this signal is challenging due to the non-perturbative production of twin gluons, but is currently under investigation~\cite{caleb}.

%%%%%%%%% CMB %%%%%%%%%%%%

The $\mathbb{Z}_2$FTH model could also give rise to a small but tell-tale cosmological signature: a single light twin neutrino that contributes to $\Delta N_\mathrm{eff}$, assuming $\nu_\tau'$ does not receive an extra contribution to its mass from an operator similar to the first term in \eref{e.newyukawa}.
The Higgs portal in this model keeps the visible and twin sectors in kinetic equilibrium in the early universe until $T \sim \mathcal{O}(\mathrm{GeV})$, at which point the two sectors decouple~\cite{Craig:2015xla}. The twin bath at that time consists of twin bottoms and twin neutrinos, with the former keeping the latter in thermal equilibrium with the SM bath until decoupling. Because the twin sector has far fewer active degrees of freedom than the visible sector, and due to subsequent entropy injections in the visible sector below the decoupling temperature, we find that the single twin neutrino only contributes 
\begin{equation}
\Delta N_\mathrm{eff}^{\mathbb{Z}_2\mathrm{FTH}} \approx 0.1
\end{equation}
to the abundance of dark radiation, in agreement with the results of Ref.~\cite{Craig:2015xla}. While this is well below current bounds~\cite{Aghanim:2018eyx}, it would be detected by upcoming Stage 4 CMB experiments, which are projected to be sensitive to $\Delta N_{\rm eff} 
\sim 0.02$~\cite{Abazajian:2016yjj}.

It has not escaped our attention that charged scalars can contribute to $\Delta a_\mu$, and thus could possibly resolve the $(g-2)_\mu$ anomaly \cite{Bennett:2006fi, Abi:2021gix}. In our scenario with $m_\phi$ above current bounds, however, this would requires a large positive value of $\xi_\mu$ and a large negative value of $\lambda_\mu$, which seems difficult to realize in a reasonable UV completion that respects flavor constraints.

Turning our attention to the collider signature of this model, the charged $Y = 2$ scalar in the visible sector, $\phi$, with coupling to taus, also gives rise to important signatures that could allow for colliders to probe this particular scenario. The high-luminosity LHC is projected to be sensitive to $m_{\phi}$ up to 1 TeV for decays to taus, and up to 2 TeV for decays to electrons or muons~\cite{Batell:2019ptb,Fuks:2013lya}. Since $m_\phi \sim \sqrt{\delta} f_\phi$, where $f_\phi \approx m_\gamma$ (see Fig.~\ref{f.darkphotonmass}) and $\delta$ represents a combination of quartics that should not be very large, this should cover much of this model's parameter space.

The existence of the massive twin photon means that a kinetic mixing with the SM hypercharge gauge boson may be present, $\frac{1}{2}\epsilon B'_{\mu \nu} B^{\mu \nu}$. Depending on the UV completion of the model, this mixing could be generated at one-loop and be large, $\epsilon \sim 0.01 - 0.1$, especially for strongly coupled UV completions. It is also entirely plausible that the kinetic mixing could be many orders of magnitude smaller. The signatures of this kinetic mixing were investigated recently in Ref.~\cite{Chacko:2019jgi}, and significant parameter space could be covered by the high-luminosity LHC and other future colliders.

Finally, the standard signatures of the FTH scenario are present in the $\mathbb{Z}_2$ scenario as well, including exotic Higgs decays into twin fermions and glueballs which give rise to long-lived particle signatures~\cite{Craig:2015pha, Curtin:2015fna, Csaki:2015fba, 
Cheng:2015buv,
Pierce:2017taw,
Lichtenstein:2018kno,
Kilic:2018sew,
Alipour-fard:2018mre,
Li:2019ulz},
the production of new SM singlet scalar states~\cite{Buttazzo:2015bka, Chacko:2017xpd, Bishara:2018sgl, Kilic:2018sew},
and possible signals of the UV completion (see e.g.~\cite{Cheng:2016uqk}).
The long-lived particle signatures are especially spectacular, and are a necessary component of our model since the annihilation of twin taus into twin bottoms relies on the $b'$ decaying or annihilating into SM particles prior to the onset of Big Bang nucleosynthesis, either directly as unstable twin bottomonia, or by annihilating into twin glueballs which decay through the Higgs portal. 
It is noteworthy that long-lived particle searches at the LHC~\cite{ATLAS:2021flf, Sirunyan:2020cao, CMS:2021uxj, Aaboud:2018aqj, Aaboud:2019opc, CMS:2021zdu}
 are starting to be sensitive to relevant parts of the FTH parameter space, in particular the production of twin bottomonia which subsequently decay to twin glueballs. While unknown aspects of the non-perturbative dynamics of the hidden sector make precise predictions challenging in some cases, it is clear that there is exciting discovery potential for Twin Higgs signatures at the LHC, with near-complete coverage of Neutral Naturalness expected with planned future experiments. We discuss this in more detail in  \aref{a.lhcllp}. 

\section{Conclusion}

In this study, we have presented a $\mathbb{Z}_2$ symmetric version of the Fraternal Twin Higgs model, which contains a viable dark matter candidate in the form of the dominantly right-handed twin tau, $\tau'_1 \approx \tau'_R$. This $\mathbb{Z}_2$FTH model extends the standard Fraternal Twin Higgs scenario by adding twin-hypercharge-breaking scalars with $Y = 2$ in order to supply the necessary $\mathbb{Z}_2$ symmetry breaking~\cite{Batell:2019ptb}. 
This addition allows us to rescue the Fraternal Twin WIMP Miracle~\cite{Garcia:2015loa,Craig:2015xla} by decoupling the twin tau mass from its Yukawa coupling to the twin Higgs, and by adding annihilation channels which proceed through twin photon exchange. We identify significant regions of parameter space in this model in which an $\mathcal{O}(100 \gev)$ twin tau can be produced thermally in the early universe with an acceptable relic abundance, while scattering with nuclei at a rate that is consistent with existing direct detection constraints. Incidentally, the truncated nature of the twin sector in fraternal models also solves the domain wall problem that is ordinarily present in Mirror Twin Higgs realizations of this spontaneous $\mathbb{Z}_2$-breaking mechanism.

This model has a number of observable consequences that will allow for its discovery or exclusion in the coming years. 
Across much of the currently viable parameter space, the dominantly-$\tau'_R$ dark matter candidate could be discovered at upcoming direct detection experiments. In addition, this scenario is predicted to generate a small but detectable contribution to the energy density of dark radiation, $\Delta N_{\rm eff} \approx 0.1$, from the single generation of light twin neutrino, well within the projected sensitivity of Stage 4 CMB experiments. 
The prospects for testing this model at colliders are also very promising.
In regions of parameter space that are compatible with $\tau'_R$ dark matter, the $Y=2$ visible-sector scalar contained in this model is expected to be within reach of searches at the high-luminosity LHC. Furthermore, the usual signals of Neutral Naturalness, such as Higgs coupling deviations and the production of long-lived particles in exotic Higgs decays, are present in this model as well. As we discuss in \aref{a.lhcllp}, the LHC is becoming sensitive to these signatures in important regions of parameter space.
Finally, indirect detection could provide an early discovery channel for this fraternal twin tau dark matter candidate, once the non-perturbative aspects of the glueball shower produced in the annihilation are understood~\cite{caleb}. 

In summary, the $\mathbb{Z}_2$FTH model presented here represents a promising and minimal extension of the Fraternal Twin Higgs framework, which addresses several of the shortcomings of the original model while preserving its solution to the little hierarchy problem. This model therefore constitutes an important new benchmark for WIMP dark matter in the Neutral Naturalness paradigm. 

\vspace{5mm}

\emph{Acknowledgements:} 
The authors would like to thank Jared Evans, Caleb Gemmell and Yuhsin Tsai for helpful discussions. 
This work was conceived at the Aspen Center for Physics, which is supported by National Science Foundation grant PHY-1607611, for which the authors are very grateful. The research of Setford, Gryba and Curtin is supported in part by a Discovery Grant from the Natural Sciences and Engineering Research Council of Canada, and by the Canada Research Chair program. Setford also acknowledges support from the University of Toronto Faculty of Arts and Science postdoctoral fellowship, and Gryba acknowledges funding from a Postgraduate Doctoral Scholarship (PGS D) provided by the Natural Sciences and Engineering Research Council of Canada. 
Hooper  is  supported  by  the  Fermi  Research  Alliance,  LLC  under Contract No.   DE-AC02-07CH11359 with the U.S. Department  of  Energy,  Office  of  High  Energy  Physics.
Scholtz is supported by the following grants: {\sc Departments of Excellence} grant awarded by the Italian Ministry of Education, University and Research ({\sc Miur}), Research grant {\sl The Dark Universe: A Synergic Multimessenger Approach}, No.~2017X7X85K funded by the Italian Ministry of Education, University and Research ({\sc Miur}), and Research grant {\sc TAsP} (Theoretical Astroparticle Physics) funded by Instituto Nazionale di Fisica Nucleare.

\vspace{5mm}

\appendix 

\section{Domain Walls}
\label{app:domainwalls}

The full Lagrangian of the scalar sector of the $\mathbb{Z}_2$FTH model can be split into two parts. First, the $U(2)$ preserving terms are given by
\begin{equation} 
\lambda_\Phi (|\phi|^2 + |\phi'|^2)^2 + \mu_\Phi^2 (|\phi|^2 + |\phi'|^2),
\end{equation}
while the $U(2)$ breaking terms are written as follows:
\begin{equation}
\delta_\Phi (|\phi|^4 + |\phi'|^4) + m^2 (|\phi|^2 - |\phi'|^2) + \delta' (|\phi|^4 - |\phi'|^4).
\end{equation}
While $m$ and $\delta'$ also explicitly break the $\mathbb{Z}_2$ symmetry, they are set to zero at tree-level in our model.

The term proportional to $\delta_\Phi$ ensures that there are only vacua of the form $(f_\phi,0)$ and $(0,f_\phi)$. It does not, however, split their energy degeneracy; only the $m^2$ and $\delta'$ terms break the degeneracy between the depth of the vacua. As a result, if $m^2 = 0$ and $\delta' =  0$, the model will predict the existence of domain walls which are disfavored by cosmology.  As long as the difference between the energy densities of the two vacua is large compared to the domain wall energy, the domain walls dissipate gravitationally before they dominate the energy density of the universe \cite{Vilenkin:1981zs}. This requirement corresponds to
\begin{equation}
\Delta V > \frac{\sigma^2}{m_{\rm Pl}^2} \ ,\label{eq:domainwalls}
\end{equation}
where $m_\mathrm{Pl}$ is the reduced Planck mass, and the domain wall tension $\sigma$ can be estimated as $\sigma^2 \approx \delta_\Phi f_\phi^6$.

The authors of Ref.~\cite{Batell:2019ptb} included a small $m\neq 0$ term in order to split the vacuum degeneracy. In our case, we get a contribution to $\Delta V$ from a nonzero $\delta'$, which is generated radiatively because the two sectors have different numbers of fermion flavors. For small $\delta'$, the energy density difference between the two vacua is 
\begin{equation}
\Delta V = \mu_\Phi^4 \frac{\delta'}{(\delta_\Phi+\lambda_\Phi)^2} \approx \delta' f_{\phi}^4.
\end{equation}
The radiative correction to $\delta'$ is generated by couplings of the first two SM lepton generations to $\phi$, analogous to \eref{e.LRR}, which are absent in the fraternal twin sector:
\begin{equation}
\delta' \sim  \sum_{l = e, \mu}  \frac{1}{16\pi^2} \lambda_l^4 \log(m_{\phi}/\Lambda).
\end{equation}
In order to satisfy condition in \eref{eq:domainwalls}, we require:
\begin{equation}
\lambda_l > \mathcal{O}(1) (f_\phi/m_{\rm Pl})^{1/2} \sim 10^{-8}.
\end{equation}
This a very small coupling, even smaller than the electron Yukawa coupling. Therefore, as long as the $\phi$ scalar has even a very tiny coupling to the light SM leptons, the vacuum in which $\phi'$ gets a twin-hypercharge-breaking VEV is preferred and the domain wall problem is resolved.

\section{Cross sections}
\label{a.xsecs}

In this appendix we present analytic expressions for the cross sections required to compute the twin tau's thermal relic abundance.
In the following expressions $Q_X, W_X, Z_X$ represent the gauge couplings of the twin photon, $W$ and $Z$ bosons, respectively, to particle $X$, where all particles are mass eigenstates. For instance $Q_{\tau'}$ represents the coupling of the twin photon mass eigenstate to the lighter $\tau_1'$ mass eigenstate ({\it i.e.}\,\,the dark matter candidate). Thus to obtain the correct couplings, both the gauge and $\tau'$ mass matrices must first be diagonalized.

The $\tau'$ mass eigenstates are
\begin{align}
  \tau_1' &= \cos\theta_{\tau'} \tau_L' - \sin\theta_{\tau'} \tau_R', \\
  \tau_2' &= \sin\theta_{\tau'} \tau_L' + \cos\theta_{\tau'} \tau_R', \nonumber
\end{align}
where the $\tau'$ mixing angle is given by
\begin{multline}
  \sin^2\theta_{\tau'} = \frac{1}{2} \\ + \frac{f_\phi(2\lambda_\tau\Lambda^2  - \xi_\tau f^2)}{2\sqrt{2f^2(\tilde c_\tau f_\phi^2 + y_\tau \Lambda^2)^2 + f_\phi^2(2 \lambda_\tau \Lambda^2 - \xi_\tau f^2)^2}}.
\end{multline}

In analogy to the SM, the twin gauge boson mass eigenstates are given by
\begin{align}
  Z_\mu' &= \cos\theta_{W'} {W_\mu^3}' - \sin\theta_{W'} B_\mu', \\
  \gamma_\mu' &= \sin\theta_{W'} {W_\mu^3}' + \cos\theta_{W'} B_\mu', \nonumber
\end{align}
where the twin Weinberg angle is given by
\begin{equation}
  \sin^2\theta_{W'} =
  \frac{(g^2-g'^2)f^2 - 32 g'^2 f_\phi^2 + \Delta}{2\Delta},
\end{equation}
with
\begin{equation}
  \Delta = \sqrt{1024 g'^4 f_\phi^4 - 64g'^2(g^2-g'^2) f_\phi^2 f^2 + (g^2+g'^2)^2 f^4}. 
\end{equation}
Keep in mind that the above definitions for $\theta_{\tau'}$ and $\theta_{W'}$ are subject to relabeling $\sin \leftrightarrow \cos$ in order to maintain $\tau'_1$ as the lightest eigenstate and the $Z'$ as the majority-$W'$ state.

\subsection{$\tau_1'\tau_1' \rightarrow \nu'\nu'$}

\begin{equation}
  \frac{d\sigma_{\tau'\tau'\rightarrow\nu'\nu'}}{d \cos\theta} = \frac{1}{64\pi^2 s} \frac{\sqrt{s}}{\sqrt{s - 4m_{\tau'}^2}} |\mathcal M_{\nu'\nu'}|^2,
\end{equation}
with
\begin{multline}
  |\mathcal M_{\nu'\nu'}|^2 =  \left(\frac{Q_{\nu'} Q_{\tau'}}{s-m_{\gamma'}^2} + \frac{Z_{\nu'} Z_{\tau'}}{s-m_{Z'}^2}\right)^2 (\alpha_+ + \alpha_-) \\ 
  + 2|W_{\tau'}|^2   \left(\frac{Q_{\nu'} Q_{\tau'}}{s-m_{\gamma'}^2} + \frac{Z_{\nu'} Z_{\tau'}}{s-m_{Z'}^2}\right) \left(\frac{\alpha_+}{t - m_{W'}^2} + \frac{\alpha_-}{u - m_{W'}^2}\right) \\
  + |W_{\tau'}|^4 \Bigg(\frac{\alpha_+}{(t-m_{W'}^2)^2} + \frac{\alpha_-}{(u-m_{W'}^2)^2} \\- \frac{8m_{\tau'}^2 s}{(t-m_{W'}^2)(u-m_{W'}^2)}\Bigg) \\
  - \frac{8 |W_{\tau'}|^2 m_{\tau'}^2 s \beta}{(s-m_{\gamma'}^2)(s-m_{Z'}^2)} \left(\frac{1}{t-m_{W'}^2} + \frac{1}{u-m_{W'}^2}\right) \\
  - \frac{8 m_{\tau'}^2 s \beta^2}{(s-m_{Z'}^2)^2 (s-m_{\gamma'}^2)^2},
\end{multline}
where
\begin{align}
  &\alpha_\pm = \left(s \pm \cos\theta \sqrt{s} \sqrt{s-4m_{\tau'}^2}\right)^2 \\
  &\beta = Q_{\nu'} Q_{\tau'} (s-m_{Z'}^2) + Z_{\nu'} Z_{\tau'} (s-m_{\gamma'}^2), \nonumber
\end{align}
and
\begin{align}
  t = - \frac{s}{4}(1 - \cos^2\theta) - \left(\frac{\cos\theta \sqrt{s}}{2} - \sqrt{\frac{s}{4} - m_{\tau'}^2} \right)^2, \nonumber \\
  u = - \frac{s}{4}(1 - \cos^2\theta) - \left(\frac{\cos\theta \sqrt{s}}{2} + \sqrt{\frac{s}{4} - m_{\tau'}^2} \right)^2. 
\end{align}

\subsection{${\tau'}_1{\tau'}_1 \rightarrow b'\bar{b}'$}

\begin{equation}
  \frac{d\sigma_{{\tau'}{\tau'}\rightarrow b'\bar{b}'}}{d \cos\theta} = \frac{1}{64\pi^2 s} \frac{\sqrt{s - 4m_{b'}^2}}{\sqrt{s - 4m_{\tau'}^2}} |\mathcal M_{b'\bar{b}'}|^2,
\end{equation}
with
\begin{multline}
  |\mathcal M_{b'\bar{b}'}|^2 = \\
   \frac{4 Q_{\tau'}^2}{(s-m_{\gamma'}^2)^2}
   \bigg( (Q_{b'}^2 + Q_{\bar{b}'}^2)\kappa + Q_{b'} Q_{\bar{b}'} m_{b'}^2 (s-6m_{\tau'}^2) \bigg) \\
   + \frac{4 Z_{\tau'}^2}{(s-m_{Z'}^2)^2}
   \bigg( (Z_{b'}^2 + Z_{\bar{b}'}^2)\kappa + Z_{b'} Z_{\bar{b}'} m_{b'}^2 (s-6m_{\tau'}^2) \bigg) \\
   + \frac{4 Q_{\tau'} Z_{\tau'}}{(s-m_{\gamma'}^2)(s-m_{Z'}^2)} \times \\
   \bigg( 2(Q_{b'} Z_{b'} + Q_{\bar{b}'} Z_{\bar{b}'})\kappa + (Q_{b'} Z_{\bar{b}'} + Q_{\bar{b}'} Z_{b'}) m_{b'}^2 (s-6m_{\tau'}^2) \bigg),
\end{multline}
where
\begin{multline}
  \kappa = \frac{1}{8} \Big(s(s-4m_{\tau'}^2)(1+\cos^2\theta) \\ + 4m_{b'}^2(2m_{\tau'}^2(1+2\cos^2\theta) - s\cos^2\theta)\Big)
\end{multline}
and
\begin{multline}
  t = - \left(\frac{s}{4} - m_{b'}^2\right)(1 - \cos^2\theta) \\- \left(\cos\theta\sqrt{\frac{s}{4} - m_{b'}^2} - \sqrt{\frac{s}{4} - m_{\tau'}^2} \right)^2,
\end{multline}
\begin{multline}
  u = - \left(\frac{s}{4} - m_{b'}^2\right)(1 - \cos^2\theta) \\- \left(\cos\theta\sqrt{\frac{s}{4} - m_{b'}^2} + \sqrt{\frac{s}{4} - m_{\tau'}^2} \right)^2.
\end{multline}

\section{Constraints on the Fraternal Twin Higgs from LHC long-lived particle searches}
\label{a.lhcllp}

In the last few years, searches for long-lived particles (LLPs) at the LHC have advanced in great strides, with both ATLAS and CMS obtaining significant sensitivity to LLPs produced in exotic higgs decays in the inner detector~\cite{ATLAS:2021flf, Sirunyan:2020cao, CMS:2021uxj} or outer detectors~\cite{Aaboud:2018aqj, Aaboud:2019opc, CMS:2021zdu}, most importantly the muon system. In this appendix, we discuss how these searches are starting to actually access the motivated parameter space of the Fraternal Twin Higgs model. 

We make a number of assumptions for simplicity and concreteness in this discussion.
First, we let the twin bottom mass obey the $\mathbb{Z}_2$ symmetry as in the $\mathbb{Z}_2$FTH scenario, $m_{b'} = (f/v) \,m_b \approx 14, 23, 32 \gev$ for $f/v = 3, 5, 7$. For illustrative purposes, we can assume that the twin bottomonium mass is simply twice the twin bottom mass (this is a reasonable approximation for the most unstable pseudoscalar state, $\chi_0$, but in reality there are different states with different binding energies~\cite{Craig:2015pha}). Similarly, we focus on just the lightest $0^{++}$ glueball with the shortest lifetime of all the glueball states~\cite{Juknevich:2009ji, Juknevich:2009gg}. RG arguments favour the glueball mass to be in the range $m_{GB} \sim 10 - 30 \gev$~\cite{Curtin:2015fna}, with a lifetime of $c \tau \sim 1 - 1000 \,\mathrm{m}$ (lighter glueballs being longer-lived), though other masses are certainly possible depending on details of the UV completion.

For this range of parameters, there are two possibilities for the cosmological history of the twin sector: either twin bottoms form bottomonia below the confinement scale, which quickly \emph{decay} to twin glueballs ($m_{GB} \lesssim m_{b'}$), or the twin bottomonia efficiently \emph{annihilate} to twin glueballs ($m_{b'} \lesssim m_{GB} \lesssim 2 m_{b'}$). Both of these scenarios are cosmologically viable, since the twin glueballs efficiently annihilate or decay to the SM~\cite{Garcia:2015loa}. 
Assuming the above mass ranges for twin glueballs, it seems that $f/v$ = 3 (7) slightly favours the twin bottomonium annihilation (twin bottomonium decay) scenario.

Both glueballs and bottomonia are produced in exotic Higgs decays at the LHC, with $\mathrm{Br}(h\to {b' b'}) \sim (0.06) \times (3 v/f)$ and $\mathrm{Br}(h \to \mathrm{twin\ glueballs}) \sim 10^{-5} - 10^{-4}$.
In the twin bottomonium decay scenario, the exotic Higgs decays to twin bottomonia would produce further twin glueballs. 
With all this in mind, we can understand how recent LLP searches at the LHC are starting to constrain the FTH scenario.

The twin bottomonium decay scenario is already probed by searches for LHC decays in the ATLAS or CMS outer detectors~\cite{Aaboud:2018aqj, Aaboud:2019opc, CMS:2021zdu}, which can probe LLP lifetimes up to the (1 - 3) $\times 10 \,\mathrm{m}$ range for percent-level exotic Higgs decay branching fractions. Assuming the bottomonia promptly decay to twin glueballs, this is  sensitive to glueball masses near $m_{b'}$~\footnote{This assumes that a sizeable fraction of produced glueballs are the shortest-lived lightest $0^{++}$ state, which is backed up by finite-temperature QCD estimates~\cite{Juknevich:2010rhj}}. By the time of the high-luminosity LHC, these searches will reach into very significant portions of FTH parameter space, though large portions are also unlikely to be excluded for $f/v \gtrsim 4$. 

Direct Higgs decays to twin glueballs can occur in both the twin bottomonium decay and annihilation scenarios, but the branching fraction is likely out of reach of main detector searches even at the high-luminosity LHC. However, MATHUSLA~\cite{Curtin:2018mvb, Alpigiani:2020tva} would cover this range completely under the above assumptions (for both direct twin glueball production or production in twin bottomonia decays). 

In the twin bottomonium annihilation scenario, LLP searches would have to detect the decays of twin bottomonia themselves. The shortest-lived pseudoscalar state, $\chi_0$, has a lifetime in the range of $c \tau \sim (10^{-4}  - 10^{-3}) \,\mathrm{m}$~\cite{Cheng:2015buv}. If all exotic Higgs decays produce $\chi_0$'s, then there is currently no sensitivity in this lifetime range for the relevant percent-level Higgs branching fractions. This is a very difficult signal to observe~\cite{Ito:2017dpm, Ito:2018asa}, even with full high-luminosity LHC data, and direct probes may require future lepton colliders (taking advantage of the clean environment and known initial state) or a 100 TeV proton-proton collider (taking advantage of the high boost of the parent Higgs bosons to increase the twin bottomonium decay length). However, in this scenario the presence of other twin bottomonium states could be important, since they can be significantly longer-lived. This could place their lifetime in the $(10^{-2} - 1)~\mathrm{m}$ window where inner detector searches have sensitivity for exotic Higgs decay production of LLPs with branching ratios of $\sim 10\%$ for for LLP masses below $\sim$ 40 GeV \cite{CMS:2021uxj,ATLAS:2021flf}, taking advantage of $Zh$ production for triggering, or displaced jet searches with pecent-level branching ratio sensitivity for LLP masses above $\sim 40 \gev$~\cite{Sirunyan:2020cao}. For longer lifetimes, the outer detector searches could have relevant sensitivity as well~\cite{Aaboud:2018aqj, Aaboud:2019opc, CMS:2021zdu}. 
While this discovery prospect is exciting, it is challenging to assess to what extent these searches actually constrain the FTH, since the details of the twin bottomonium spectrum and the lifetime of the higher states are highly uncertain~\cite{Craig:2015pha, Cheng:2015buv}. 

While LHC LLP searches are already starting to probe important regions of FTH parameter space, there are some regions in which our ability to interpret these bounds is hindered by unknown aspects of the twin sector's non-perturbative dynamics. Future detectors like MATHUSLA would allow for near-complete coverage of the most motivated range of signals with lifetimes above a meter, while future lepton and proton colliders would allow the entire FTH model space to be probed, by a combination of LLP searches~\cite{Liu:2016zki,
Alipour-Fard:2018lsf,
Cheung:2019qdr, Curtin:2015fna}, scalar resonance searches~\cite{Buttazzo:2015bka, Chacko:2017xpd, Bishara:2018sgl, Kilic:2018sew}, and Higgs coupling measurements \cite{Burdman:2014zta,Dawson:2013bba, Craig:2013xia} that are sensitive to the $\cos \vartheta$ reduction of the Higgs couplings to SM fermions regardless of the detailed twin sector phenomenology.

\bibliography{References}

%merlin.mbs apsrev4-1.bst 2010-07-25 4.21a (PWD, AO, DPC) hacked
%Control: key (0)
%Control: author (8) initials jnrlst
%Control: editor formatted (1) identically to author
%Control: production of article title (-1) disabled
%Control: page (0) single
%Control: year (1) truncated
%Control: production of eprint (0) enabled
\begin{thebibliography}{112}%
\makeatletter
\providecommand \@ifxundefined [1]{%
 \@ifx{#1\undefined}
}%
\providecommand \@ifnum [1]{%
 \ifnum #1\expandafter \@firstoftwo
 \else \expandafter \@secondoftwo
 \fi
}%
\providecommand \@ifx [1]{%
 \ifx #1\expandafter \@firstoftwo
 \else \expandafter \@secondoftwo
 \fi
}%
\providecommand \natexlab [1]{#1}%
\providecommand \enquote  [1]{``#1''}%
\providecommand \bibnamefont  [1]{#1}%
\providecommand \bibfnamefont [1]{#1}%
\providecommand \citenamefont [1]{#1}%
\providecommand \href@noop [0]{\@secondoftwo}%
\providecommand \href [0]{\begingroup \@sanitize@url \@href}%
\providecommand \@href[1]{\@@startlink{#1}\@@href}%
\providecommand \@@href[1]{\endgroup#1\@@endlink}%
\providecommand \@sanitize@url [0]{\catcode `\\12\catcode `\$12\catcode
  `\&12\catcode `\#12\catcode `\^12\catcode `\_12\catcode `\%12\relax}%
\providecommand \@@startlink[1]{}%
\providecommand \@@endlink[0]{}%
\providecommand \url  [0]{\begingroup\@sanitize@url \@url }%
\providecommand \@url [1]{\endgroup\@href {#1}{\urlprefix }}%
\providecommand \urlprefix  [0]{URL }%
\providecommand \Eprint [0]{\href }%
\providecommand \doibase [0]{http://dx.doi.org/}%
\providecommand \selectlanguage [0]{\@gobble}%
\providecommand \bibinfo  [0]{\@secondoftwo}%
\providecommand \bibfield  [0]{\@secondoftwo}%
\providecommand \translation [1]{[#1]}%
\providecommand \BibitemOpen [0]{}%
\providecommand \bibitemStop [0]{}%
\providecommand \bibitemNoStop [0]{.\EOS\space}%
\providecommand \EOS [0]{\spacefactor3000\relax}%
\providecommand \BibitemShut  [1]{\csname bibitem#1\endcsname}%
\let\auto@bib@innerbib\@empty
%</preamble>
\bibitem [{\citenamefont {Bertone}\ and\ \citenamefont
  {Hooper}(2016)}]{Bertone:2016nfn}%
  \BibitemOpen
  \bibfield  {author} {\bibinfo {author} {\bibfnamefont {G.}~\bibnamefont
  {Bertone}}\ and\ \bibinfo {author} {\bibfnamefont {D.}~\bibnamefont
  {Hooper}},\ }\href@noop {} {\bibfield  {journal} {\bibinfo  {journal}
  {Submitted to: Rev. Mod. Phys.}\ } (\bibinfo {year} {2016})},\ \Eprint
  {http://arxiv.org/abs/1605.04909} {arXiv:1605.04909 [astro-ph.CO]}
  \BibitemShut {NoStop}%
%%CITATION = ARXIV:1605.04909;%%
\bibitem [{\citenamefont {Martin}(2010)}]{Martin:1997ns}%
  \BibitemOpen
  \bibfield  {author} {\bibinfo {author} {\bibfnamefont {S.~P.}\ \bibnamefont
  {Martin}},\ }\href {\doibase 10.1142/9789814307505_0001} {\bibfield
  {journal} {\bibinfo  {journal} {Adv.Ser.Direct.High Energy Phys.}\ }\textbf
  {\bibinfo {volume} {21}},\ \bibinfo {pages} {1} (\bibinfo {year} {2010})},\
  \Eprint {http://arxiv.org/abs/hep-ph/9709356} {arXiv:hep-ph/9709356 [hep-ph]}
  \BibitemShut {NoStop}%
%%CITATION = HEP-PH/9709356;%%
\bibitem [{\citenamefont {Aaboud}\ \emph {et~al.}(2017)\citenamefont {Aaboud}
  \emph {et~al.}}]{Aaboud:2017ayj}%
  \BibitemOpen
  \bibfield  {author} {\bibinfo {author} {\bibfnamefont {M.}~\bibnamefont
  {Aaboud}} \emph {et~al.} (\bibinfo {collaboration} {ATLAS}),\ }\href
  {\doibase 10.1007/JHEP12(2017)085} {\bibfield  {journal} {\bibinfo  {journal}
  {JHEP}\ }\textbf {\bibinfo {volume} {12}},\ \bibinfo {pages} {085} (\bibinfo
  {year} {2017})},\ \Eprint {http://arxiv.org/abs/1709.04183} {arXiv:1709.04183
  [hep-ex]} \BibitemShut {NoStop}%
%%CITATION = ARXIV:1709.04183;%%
\bibitem [{\citenamefont {Aaboud}\ \emph
  {et~al.}(2018{\natexlab{a}})\citenamefont {Aaboud} \emph
  {et~al.}}]{Aaboud:2017aeu}%
  \BibitemOpen
  \bibfield  {author} {\bibinfo {author} {\bibfnamefont {M.}~\bibnamefont
  {Aaboud}} \emph {et~al.} (\bibinfo {collaboration} {ATLAS}),\ }\href
  {\doibase 10.1007/JHEP06(2018)108} {\bibfield  {journal} {\bibinfo  {journal}
  {JHEP}\ }\textbf {\bibinfo {volume} {06}},\ \bibinfo {pages} {108} (\bibinfo
  {year} {2018}{\natexlab{a}})},\ \Eprint {http://arxiv.org/abs/1711.11520}
  {arXiv:1711.11520 [hep-ex]} \BibitemShut {NoStop}%
%%CITATION = ARXIV:1711.11520;%%
\bibitem [{\citenamefont {Aaboud}\ \emph
  {et~al.}(2018{\natexlab{b}})\citenamefont {Aaboud} \emph
  {et~al.}}]{Aaboud:2018zjf}%
  \BibitemOpen
  \bibfield  {author} {\bibinfo {author} {\bibfnamefont {M.}~\bibnamefont
  {Aaboud}} \emph {et~al.} (\bibinfo {collaboration} {ATLAS}),\ }\href
  {\doibase 10.1007/JHEP09(2018)050} {\bibfield  {journal} {\bibinfo  {journal}
  {JHEP}\ }\textbf {\bibinfo {volume} {09}},\ \bibinfo {pages} {050} (\bibinfo
  {year} {2018}{\natexlab{b}})},\ \Eprint {http://arxiv.org/abs/1805.01649}
  {arXiv:1805.01649 [hep-ex]} \BibitemShut {NoStop}%
%%CITATION = ARXIV:1805.01649;%%
\bibitem [{\citenamefont {Sirunyan}\ \emph
  {et~al.}(2018{\natexlab{a}})\citenamefont {Sirunyan} \emph
  {et~al.}}]{Sirunyan:2018lul}%
  \BibitemOpen
  \bibfield  {author} {\bibinfo {author} {\bibfnamefont {A.~M.}\ \bibnamefont
  {Sirunyan}} \emph {et~al.} (\bibinfo {collaboration} {CMS}),\ }\href
  {\doibase 10.1007/JHEP11(2018)079} {\bibfield  {journal} {\bibinfo  {journal}
  {JHEP}\ }\textbf {\bibinfo {volume} {11}},\ \bibinfo {pages} {079} (\bibinfo
  {year} {2018}{\natexlab{a}})},\ \Eprint {http://arxiv.org/abs/1807.07799}
  {arXiv:1807.07799 [hep-ex]} \BibitemShut {NoStop}%
%%CITATION = ARXIV:1807.07799;%%
\bibitem [{\citenamefont {Sirunyan}\ \emph
  {et~al.}(2018{\natexlab{b}})\citenamefont {Sirunyan} \emph
  {et~al.}}]{Sirunyan:2017leh}%
  \BibitemOpen
  \bibfield  {author} {\bibinfo {author} {\bibfnamefont {A.~M.}\ \bibnamefont
  {Sirunyan}} \emph {et~al.} (\bibinfo {collaboration} {CMS}),\ }\href
  {\doibase 10.1103/PhysRevD.97.032009} {\bibfield  {journal} {\bibinfo
  {journal} {Phys. Rev.}\ }\textbf {\bibinfo {volume} {D97}},\ \bibinfo {pages}
  {032009} (\bibinfo {year} {2018}{\natexlab{b}})},\ \Eprint
  {http://arxiv.org/abs/1711.00752} {arXiv:1711.00752 [hep-ex]} \BibitemShut
  {NoStop}%
%%CITATION = ARXIV:1711.00752;%%
\bibitem [{\citenamefont {Sirunyan}\ \emph
  {et~al.}(2018{\natexlab{c}})\citenamefont {Sirunyan} \emph
  {et~al.}}]{Sirunyan:2017kiw}%
  \BibitemOpen
  \bibfield  {author} {\bibinfo {author} {\bibfnamefont {A.~M.}\ \bibnamefont
  {Sirunyan}} \emph {et~al.} (\bibinfo {collaboration} {CMS}),\ }\href
  {\doibase 10.1016/j.physletb.2018.01.012} {\bibfield  {journal} {\bibinfo
  {journal} {Phys. Lett.}\ }\textbf {\bibinfo {volume} {B778}},\ \bibinfo
  {pages} {263} (\bibinfo {year} {2018}{\natexlab{c}})},\ \Eprint
  {http://arxiv.org/abs/1707.07274} {arXiv:1707.07274 [hep-ex]} \BibitemShut
  {NoStop}%
%%CITATION = ARXIV:1707.07274;%%
\bibitem [{\citenamefont {Aprile}\ \emph {et~al.}(2018)\citenamefont {Aprile}
  \emph {et~al.}}]{Aprile:2018dbl}%
  \BibitemOpen
  \bibfield  {author} {\bibinfo {author} {\bibfnamefont {E.}~\bibnamefont
  {Aprile}} \emph {et~al.} (\bibinfo {collaboration} {XENON}),\ }\href
  {\doibase 10.1103/PhysRevLett.121.111302} {\bibfield  {journal} {\bibinfo
  {journal} {Phys. Rev. Lett.}\ }\textbf {\bibinfo {volume} {121}},\ \bibinfo
  {pages} {111302} (\bibinfo {year} {2018})},\ \Eprint
  {http://arxiv.org/abs/1805.12562} {arXiv:1805.12562 [astro-ph.CO]}
  \BibitemShut {NoStop}%
%%CITATION = ARXIV:1805.12562;%%
\bibitem [{\citenamefont {Akerib}\ \emph {et~al.}(2017)\citenamefont {Akerib}
  \emph {et~al.}}]{Akerib:2016vxi}%
  \BibitemOpen
  \bibfield  {author} {\bibinfo {author} {\bibfnamefont {D.~S.}\ \bibnamefont
  {Akerib}} \emph {et~al.} (\bibinfo {collaboration} {LUX}),\ }\href {\doibase
  10.1103/PhysRevLett.118.021303} {\bibfield  {journal} {\bibinfo  {journal}
  {Phys. Rev. Lett.}\ }\textbf {\bibinfo {volume} {118}},\ \bibinfo {pages}
  {021303} (\bibinfo {year} {2017})},\ \Eprint
  {http://arxiv.org/abs/1608.07648} {arXiv:1608.07648 [astro-ph.CO]}
  \BibitemShut {NoStop}%
%%CITATION = ARXIV:1608.07648;%%
\bibitem [{\citenamefont {Cui}\ \emph {et~al.}(2017{\natexlab{a}})\citenamefont
  {Cui} \emph {et~al.}}]{Cui:2017nnn}%
  \BibitemOpen
  \bibfield  {author} {\bibinfo {author} {\bibfnamefont {X.}~\bibnamefont
  {Cui}} \emph {et~al.} (\bibinfo {collaboration} {PandaX-II}),\ }\href
  {\doibase 10.1103/PhysRevLett.119.181302} {\bibfield  {journal} {\bibinfo
  {journal} {Phys. Rev. Lett.}\ }\textbf {\bibinfo {volume} {119}},\ \bibinfo
  {pages} {181302} (\bibinfo {year} {2017}{\natexlab{a}})},\ \Eprint
  {http://arxiv.org/abs/1708.06917} {arXiv:1708.06917 [astro-ph.CO]}
  \BibitemShut {NoStop}%
%%CITATION = ARXIV:1708.06917;%%
\bibitem [{\citenamefont {Chacko}\ \emph
  {et~al.}(2006{\natexlab{a}})\citenamefont {Chacko}, \citenamefont {Goh},\
  and\ \citenamefont {Harnik}}]{Chacko:2005pe}%
  \BibitemOpen
  \bibfield  {author} {\bibinfo {author} {\bibfnamefont {Z.}~\bibnamefont
  {Chacko}}, \bibinfo {author} {\bibfnamefont {H.-S.}\ \bibnamefont {Goh}}, \
  and\ \bibinfo {author} {\bibfnamefont {R.}~\bibnamefont {Harnik}},\ }\href
  {\doibase 10.1103/PhysRevLett.96.231802} {\bibfield  {journal} {\bibinfo
  {journal} {Phys. Rev. Lett.}\ }\textbf {\bibinfo {volume} {96}},\ \bibinfo
  {pages} {231802} (\bibinfo {year} {2006}{\natexlab{a}})},\ \Eprint
  {http://arxiv.org/abs/hep-ph/0506256} {arXiv:hep-ph/0506256 [hep-ph]}
  \BibitemShut {NoStop}%
%%CITATION = HEP-PH/0506256;%%
\bibitem [{\citenamefont {Barbieri}\ \emph {et~al.}(2005)\citenamefont
  {Barbieri}, \citenamefont {Gregoire},\ and\ \citenamefont
  {Hall}}]{Barbieri:2005ri}%
  \BibitemOpen
  \bibfield  {author} {\bibinfo {author} {\bibfnamefont {R.}~\bibnamefont
  {Barbieri}}, \bibinfo {author} {\bibfnamefont {T.}~\bibnamefont {Gregoire}},
  \ and\ \bibinfo {author} {\bibfnamefont {L.~J.}\ \bibnamefont {Hall}},\
  }\href@noop {} {\  (\bibinfo {year} {2005})},\ \Eprint
  {http://arxiv.org/abs/hep-ph/0509242} {arXiv:hep-ph/0509242 [hep-ph]}
  \BibitemShut {NoStop}%
%%CITATION = HEP-PH/0509242;%%
\bibitem [{\citenamefont {Chacko}\ \emph
  {et~al.}(2006{\natexlab{b}})\citenamefont {Chacko}, \citenamefont {Nomura},
  \citenamefont {Papucci},\ and\ \citenamefont {Perez}}]{Chacko:2005vw}%
  \BibitemOpen
  \bibfield  {author} {\bibinfo {author} {\bibfnamefont {Z.}~\bibnamefont
  {Chacko}}, \bibinfo {author} {\bibfnamefont {Y.}~\bibnamefont {Nomura}},
  \bibinfo {author} {\bibfnamefont {M.}~\bibnamefont {Papucci}}, \ and\
  \bibinfo {author} {\bibfnamefont {G.}~\bibnamefont {Perez}},\ }\href
  {\doibase 10.1088/1126-6708/2006/01/126} {\bibfield  {journal} {\bibinfo
  {journal} {JHEP}\ }\textbf {\bibinfo {volume} {01}},\ \bibinfo {pages} {126}
  (\bibinfo {year} {2006}{\natexlab{b}})},\ \Eprint
  {http://arxiv.org/abs/hep-ph/0510273} {arXiv:hep-ph/0510273 [hep-ph]}
  \BibitemShut {NoStop}%
%%CITATION = HEP-PH/0510273;%%
\bibitem [{\citenamefont {Burdman}\ \emph {et~al.}(2007)\citenamefont
  {Burdman}, \citenamefont {Chacko}, \citenamefont {Goh},\ and\ \citenamefont
  {Harnik}}]{Burdman:2006tz}%
  \BibitemOpen
  \bibfield  {author} {\bibinfo {author} {\bibfnamefont {G.}~\bibnamefont
  {Burdman}}, \bibinfo {author} {\bibfnamefont {Z.}~\bibnamefont {Chacko}},
  \bibinfo {author} {\bibfnamefont {H.-S.}\ \bibnamefont {Goh}}, \ and\
  \bibinfo {author} {\bibfnamefont {R.}~\bibnamefont {Harnik}},\ }\href
  {\doibase 10.1088/1126-6708/2007/02/009} {\bibfield  {journal} {\bibinfo
  {journal} {JHEP}\ }\textbf {\bibinfo {volume} {02}},\ \bibinfo {pages} {009}
  (\bibinfo {year} {2007})},\ \Eprint {http://arxiv.org/abs/hep-ph/0609152}
  {arXiv:hep-ph/0609152 [hep-ph]} \BibitemShut {NoStop}%
%%CITATION = HEP-PH/0609152;%%
\bibitem [{\citenamefont {Cai}\ \emph {et~al.}(2009)\citenamefont {Cai},
  \citenamefont {Cheng},\ and\ \citenamefont {Terning}}]{Cai:2008au}%
  \BibitemOpen
  \bibfield  {author} {\bibinfo {author} {\bibfnamefont {H.}~\bibnamefont
  {Cai}}, \bibinfo {author} {\bibfnamefont {H.-C.}\ \bibnamefont {Cheng}}, \
  and\ \bibinfo {author} {\bibfnamefont {J.}~\bibnamefont {Terning}},\ }\href
  {\doibase 10.1088/1126-6708/2009/05/045} {\bibfield  {journal} {\bibinfo
  {journal} {JHEP}\ }\textbf {\bibinfo {volume} {05}},\ \bibinfo {pages} {045}
  (\bibinfo {year} {2009})},\ \Eprint {http://arxiv.org/abs/0812.0843}
  {arXiv:0812.0843 [hep-ph]} \BibitemShut {NoStop}%
%%CITATION = ARXIV:0812.0843;%%
\bibitem [{\citenamefont {Poland}\ and\ \citenamefont
  {Thaler}(2008)}]{Poland:2008ev}%
  \BibitemOpen
  \bibfield  {author} {\bibinfo {author} {\bibfnamefont {D.}~\bibnamefont
  {Poland}}\ and\ \bibinfo {author} {\bibfnamefont {J.}~\bibnamefont
  {Thaler}},\ }\href {\doibase 10.1088/1126-6708/2008/11/083} {\bibfield
  {journal} {\bibinfo  {journal} {JHEP}\ }\textbf {\bibinfo {volume} {11}},\
  \bibinfo {pages} {083} (\bibinfo {year} {2008})},\ \Eprint
  {http://arxiv.org/abs/0808.1290} {arXiv:0808.1290 [hep-ph]} \BibitemShut
  {NoStop}%
%%CITATION = ARXIV:0808.1290;%%
\bibitem [{\citenamefont {Craig}\ \emph
  {et~al.}(2015{\natexlab{a}})\citenamefont {Craig}, \citenamefont {Katz},
  \citenamefont {Strassler},\ and\ \citenamefont {Sundrum}}]{Craig:2015pha}%
  \BibitemOpen
  \bibfield  {author} {\bibinfo {author} {\bibfnamefont {N.}~\bibnamefont
  {Craig}}, \bibinfo {author} {\bibfnamefont {A.}~\bibnamefont {Katz}},
  \bibinfo {author} {\bibfnamefont {M.}~\bibnamefont {Strassler}}, \ and\
  \bibinfo {author} {\bibfnamefont {R.}~\bibnamefont {Sundrum}},\ }\href
  {\doibase 10.1007/JHEP07(2015)105} {\bibfield  {journal} {\bibinfo  {journal}
  {JHEP}\ }\textbf {\bibinfo {volume} {07}},\ \bibinfo {pages} {105} (\bibinfo
  {year} {2015}{\natexlab{a}})},\ \Eprint {http://arxiv.org/abs/1501.05310}
  {arXiv:1501.05310 [hep-ph]} \BibitemShut {NoStop}%
%%CITATION = ARXIV:1501.05310;%%
\bibitem [{\citenamefont {Cohen}\ \emph {et~al.}(2018)\citenamefont {Cohen},
  \citenamefont {Craig}, \citenamefont {Giudice},\ and\ \citenamefont
  {Mccullough}}]{Cohen:2018mgv}%
  \BibitemOpen
  \bibfield  {author} {\bibinfo {author} {\bibfnamefont {T.}~\bibnamefont
  {Cohen}}, \bibinfo {author} {\bibfnamefont {N.}~\bibnamefont {Craig}},
  \bibinfo {author} {\bibfnamefont {G.~F.}\ \bibnamefont {Giudice}}, \ and\
  \bibinfo {author} {\bibfnamefont {M.}~\bibnamefont {Mccullough}},\ }\href
  {\doibase 10.1007/JHEP05(2018)091} {\bibfield  {journal} {\bibinfo  {journal}
  {JHEP}\ }\textbf {\bibinfo {volume} {05}},\ \bibinfo {pages} {091} (\bibinfo
  {year} {2018})},\ \Eprint {http://arxiv.org/abs/1803.03647} {arXiv:1803.03647
  [hep-ph]} \BibitemShut {NoStop}%
\bibitem [{\citenamefont {Cheng}\ \emph
  {et~al.}(2018{\natexlab{a}})\citenamefont {Cheng}, \citenamefont {Li},
  \citenamefont {Salvioni},\ and\ \citenamefont {Verhaaren}}]{Cheng:2018gvu}%
  \BibitemOpen
  \bibfield  {author} {\bibinfo {author} {\bibfnamefont {H.-C.}\ \bibnamefont
  {Cheng}}, \bibinfo {author} {\bibfnamefont {L.}~\bibnamefont {Li}}, \bibinfo
  {author} {\bibfnamefont {E.}~\bibnamefont {Salvioni}}, \ and\ \bibinfo
  {author} {\bibfnamefont {C.~B.}\ \bibnamefont {Verhaaren}},\ }\href {\doibase
  10.1007/JHEP05(2018)057} {\bibfield  {journal} {\bibinfo  {journal} {JHEP}\
  }\textbf {\bibinfo {volume} {05}},\ \bibinfo {pages} {057} (\bibinfo {year}
  {2018}{\natexlab{a}})},\ \Eprint {http://arxiv.org/abs/1803.03651}
  {arXiv:1803.03651 [hep-ph]} \BibitemShut {NoStop}%
\bibitem [{\citenamefont {Chacko}\ \emph
  {et~al.}(2006{\natexlab{c}})\citenamefont {Chacko}, \citenamefont {Goh},\
  and\ \citenamefont {Harnik}}]{Chacko:2005un}%
  \BibitemOpen
  \bibfield  {author} {\bibinfo {author} {\bibfnamefont {Z.}~\bibnamefont
  {Chacko}}, \bibinfo {author} {\bibfnamefont {H.-S.}\ \bibnamefont {Goh}}, \
  and\ \bibinfo {author} {\bibfnamefont {R.}~\bibnamefont {Harnik}},\ }\href
  {\doibase 10.1088/1126-6708/2006/01/108} {\bibfield  {journal} {\bibinfo
  {journal} {JHEP}\ }\textbf {\bibinfo {volume} {01}},\ \bibinfo {pages} {108}
  (\bibinfo {year} {2006}{\natexlab{c}})},\ \Eprint
  {http://arxiv.org/abs/hep-ph/0512088} {arXiv:hep-ph/0512088 [hep-ph]}
  \BibitemShut {NoStop}%
%%CITATION = HEP-PH/0512088;%%
\bibitem [{\citenamefont {Perelstein}(2007)}]{Perelstein:2005ka}%
  \BibitemOpen
  \bibfield  {author} {\bibinfo {author} {\bibfnamefont {M.}~\bibnamefont
  {Perelstein}},\ }\href {\doibase 10.1016/j.ppnp.2006.04.001} {\bibfield
  {journal} {\bibinfo  {journal} {Prog. Part. Nucl. Phys.}\ }\textbf {\bibinfo
  {volume} {58}},\ \bibinfo {pages} {247} (\bibinfo {year} {2007})},\ \Eprint
  {http://arxiv.org/abs/hep-ph/0512128} {arXiv:hep-ph/0512128 [hep-ph]}
  \BibitemShut {NoStop}%
%%CITATION = HEP-PH/0512128;%%
\bibitem [{\citenamefont {Craig}\ and\ \citenamefont
  {Howe}(2014)}]{Craig:2013fga}%
  \BibitemOpen
  \bibfield  {author} {\bibinfo {author} {\bibfnamefont {N.}~\bibnamefont
  {Craig}}\ and\ \bibinfo {author} {\bibfnamefont {K.}~\bibnamefont {Howe}},\
  }\href {\doibase 10.1007/JHEP03(2014)140} {\bibfield  {journal} {\bibinfo
  {journal} {JHEP}\ }\textbf {\bibinfo {volume} {03}},\ \bibinfo {pages} {140}
  (\bibinfo {year} {2014})},\ \Eprint {http://arxiv.org/abs/1312.1341}
  {arXiv:1312.1341 [hep-ph]} \BibitemShut {NoStop}%
\bibitem [{\citenamefont {Falkowski}\ \emph {et~al.}(2006)\citenamefont
  {Falkowski}, \citenamefont {Pokorski},\ and\ \citenamefont
  {Schmaltz}}]{Falkowski:2006qq}%
  \BibitemOpen
  \bibfield  {author} {\bibinfo {author} {\bibfnamefont {A.}~\bibnamefont
  {Falkowski}}, \bibinfo {author} {\bibfnamefont {S.}~\bibnamefont {Pokorski}},
  \ and\ \bibinfo {author} {\bibfnamefont {M.}~\bibnamefont {Schmaltz}},\
  }\href {\doibase 10.1103/PhysRevD.74.035003} {\bibfield  {journal} {\bibinfo
  {journal} {Phys. Rev. D}\ }\textbf {\bibinfo {volume} {74}},\ \bibinfo
  {pages} {035003} (\bibinfo {year} {2006})},\ \Eprint
  {http://arxiv.org/abs/hep-ph/0604066} {arXiv:hep-ph/0604066} \BibitemShut
  {NoStop}%
\bibitem [{\citenamefont {Chang}\ \emph {et~al.}(2007)\citenamefont {Chang},
  \citenamefont {Hall},\ and\ \citenamefont {Weiner}}]{Chang:2006ra}%
  \BibitemOpen
  \bibfield  {author} {\bibinfo {author} {\bibfnamefont {S.}~\bibnamefont
  {Chang}}, \bibinfo {author} {\bibfnamefont {L.~J.}\ \bibnamefont {Hall}}, \
  and\ \bibinfo {author} {\bibfnamefont {N.}~\bibnamefont {Weiner}},\ }\href
  {\doibase 10.1103/PhysRevD.75.035009} {\bibfield  {journal} {\bibinfo
  {journal} {Phys. Rev. D}\ }\textbf {\bibinfo {volume} {75}},\ \bibinfo
  {pages} {035009} (\bibinfo {year} {2007})},\ \Eprint
  {http://arxiv.org/abs/hep-ph/0604076} {arXiv:hep-ph/0604076} \BibitemShut
  {NoStop}%
\bibitem [{\citenamefont {Craig}\ \emph {et~al.}(2016)\citenamefont {Craig},
  \citenamefont {Knapen}, \citenamefont {Longhi},\ and\ \citenamefont
  {Strassler}}]{Craig:2016kue}%
  \BibitemOpen
  \bibfield  {author} {\bibinfo {author} {\bibfnamefont {N.}~\bibnamefont
  {Craig}}, \bibinfo {author} {\bibfnamefont {S.}~\bibnamefont {Knapen}},
  \bibinfo {author} {\bibfnamefont {P.}~\bibnamefont {Longhi}}, \ and\ \bibinfo
  {author} {\bibfnamefont {M.}~\bibnamefont {Strassler}},\ }\href {\doibase
  10.1007/JHEP07(2016)002} {\bibfield  {journal} {\bibinfo  {journal} {JHEP}\
  }\textbf {\bibinfo {volume} {07}},\ \bibinfo {pages} {002} (\bibinfo {year}
  {2016})},\ \Eprint {http://arxiv.org/abs/1601.07181} {arXiv:1601.07181
  [hep-ph]} \BibitemShut {NoStop}%
%%CITATION = ARXIV:1601.07181;%%
\bibitem [{\citenamefont {Katz}\ \emph {et~al.}(2017)\citenamefont {Katz},
  \citenamefont {Mariotti}, \citenamefont {Pokorski}, \citenamefont
  {Redigolo},\ and\ \citenamefont {Ziegler}}]{Katz:2016wtw}%
  \BibitemOpen
  \bibfield  {author} {\bibinfo {author} {\bibfnamefont {A.}~\bibnamefont
  {Katz}}, \bibinfo {author} {\bibfnamefont {A.}~\bibnamefont {Mariotti}},
  \bibinfo {author} {\bibfnamefont {S.}~\bibnamefont {Pokorski}}, \bibinfo
  {author} {\bibfnamefont {D.}~\bibnamefont {Redigolo}}, \ and\ \bibinfo
  {author} {\bibfnamefont {R.}~\bibnamefont {Ziegler}},\ }\href {\doibase
  10.1007/JHEP01(2017)142} {\bibfield  {journal} {\bibinfo  {journal} {JHEP}\
  }\textbf {\bibinfo {volume} {01}},\ \bibinfo {pages} {142} (\bibinfo {year}
  {2017})},\ \Eprint {http://arxiv.org/abs/1611.08615} {arXiv:1611.08615
  [hep-ph]} \BibitemShut {NoStop}%
\bibitem [{\citenamefont {Badziak}\ and\ \citenamefont
  {Harigaya}(2017{\natexlab{a}})}]{Badziak:2017syq}%
  \BibitemOpen
  \bibfield  {author} {\bibinfo {author} {\bibfnamefont {M.}~\bibnamefont
  {Badziak}}\ and\ \bibinfo {author} {\bibfnamefont {K.}~\bibnamefont
  {Harigaya}},\ }\href {\doibase 10.1007/JHEP06(2017)065} {\bibfield  {journal}
  {\bibinfo  {journal} {JHEP}\ }\textbf {\bibinfo {volume} {06}},\ \bibinfo
  {pages} {065} (\bibinfo {year} {2017}{\natexlab{a}})},\ \Eprint
  {http://arxiv.org/abs/1703.02122} {arXiv:1703.02122 [hep-ph]} \BibitemShut
  {NoStop}%
\bibitem [{\citenamefont {Badziak}\ and\ \citenamefont
  {Harigaya}(2017{\natexlab{b}})}]{Badziak:2017kjk}%
  \BibitemOpen
  \bibfield  {author} {\bibinfo {author} {\bibfnamefont {M.}~\bibnamefont
  {Badziak}}\ and\ \bibinfo {author} {\bibfnamefont {K.}~\bibnamefont
  {Harigaya}},\ }\href {\doibase 10.1007/JHEP10(2017)109} {\bibfield  {journal}
  {\bibinfo  {journal} {JHEP}\ }\textbf {\bibinfo {volume} {10}},\ \bibinfo
  {pages} {109} (\bibinfo {year} {2017}{\natexlab{b}})},\ \Eprint
  {http://arxiv.org/abs/1707.09071} {arXiv:1707.09071 [hep-ph]} \BibitemShut
  {NoStop}%
\bibitem [{\citenamefont {Badziak}\ and\ \citenamefont
  {Harigaya}(2018)}]{Badziak:2017wxn}%
  \BibitemOpen
  \bibfield  {author} {\bibinfo {author} {\bibfnamefont {M.}~\bibnamefont
  {Badziak}}\ and\ \bibinfo {author} {\bibfnamefont {K.}~\bibnamefont
  {Harigaya}},\ }\href {\doibase 10.1103/PhysRevLett.120.211803} {\bibfield
  {journal} {\bibinfo  {journal} {Phys. Rev. Lett.}\ }\textbf {\bibinfo
  {volume} {120}},\ \bibinfo {pages} {211803} (\bibinfo {year} {2018})},\
  \Eprint {http://arxiv.org/abs/1711.11040} {arXiv:1711.11040 [hep-ph]}
  \BibitemShut {NoStop}%
\bibitem [{\citenamefont {Geller}\ and\ \citenamefont
  {Telem}(2015)}]{Geller:2014kta}%
  \BibitemOpen
  \bibfield  {author} {\bibinfo {author} {\bibfnamefont {M.}~\bibnamefont
  {Geller}}\ and\ \bibinfo {author} {\bibfnamefont {O.}~\bibnamefont {Telem}},\
  }\href {\doibase 10.1103/PhysRevLett.114.191801} {\bibfield  {journal}
  {\bibinfo  {journal} {Phys. Rev. Lett.}\ }\textbf {\bibinfo {volume} {114}},\
  \bibinfo {pages} {191801} (\bibinfo {year} {2015})},\ \Eprint
  {http://arxiv.org/abs/1411.2974} {arXiv:1411.2974 [hep-ph]} \BibitemShut
  {NoStop}%
\bibitem [{\citenamefont {Barbieri}\ \emph {et~al.}(2015)\citenamefont
  {Barbieri}, \citenamefont {Greco}, \citenamefont {Rattazzi},\ and\
  \citenamefont {Wulzer}}]{Barbieri:2015lqa}%
  \BibitemOpen
  \bibfield  {author} {\bibinfo {author} {\bibfnamefont {R.}~\bibnamefont
  {Barbieri}}, \bibinfo {author} {\bibfnamefont {D.}~\bibnamefont {Greco}},
  \bibinfo {author} {\bibfnamefont {R.}~\bibnamefont {Rattazzi}}, \ and\
  \bibinfo {author} {\bibfnamefont {A.}~\bibnamefont {Wulzer}},\ }\href
  {\doibase 10.1007/JHEP08(2015)161} {\bibfield  {journal} {\bibinfo  {journal}
  {JHEP}\ }\textbf {\bibinfo {volume} {08}},\ \bibinfo {pages} {161} (\bibinfo
  {year} {2015})},\ \Eprint {http://arxiv.org/abs/1501.07803} {arXiv:1501.07803
  [hep-ph]} \BibitemShut {NoStop}%
\bibitem [{\citenamefont {Low}\ \emph {et~al.}(2015)\citenamefont {Low},
  \citenamefont {Tesi},\ and\ \citenamefont {Wang}}]{Low:2015nqa}%
  \BibitemOpen
  \bibfield  {author} {\bibinfo {author} {\bibfnamefont {M.}~\bibnamefont
  {Low}}, \bibinfo {author} {\bibfnamefont {A.}~\bibnamefont {Tesi}}, \ and\
  \bibinfo {author} {\bibfnamefont {L.-T.}\ \bibnamefont {Wang}},\ }\href
  {\doibase 10.1103/PhysRevD.91.095012} {\bibfield  {journal} {\bibinfo
  {journal} {Phys. Rev. D}\ }\textbf {\bibinfo {volume} {91}},\ \bibinfo
  {pages} {095012} (\bibinfo {year} {2015})},\ \Eprint
  {http://arxiv.org/abs/1501.07890} {arXiv:1501.07890 [hep-ph]} \BibitemShut
  {NoStop}%
\bibitem [{\citenamefont {Burdman}\ \emph {et~al.}(2015)\citenamefont
  {Burdman}, \citenamefont {Chacko}, \citenamefont {Harnik}, \citenamefont
  {de~Lima},\ and\ \citenamefont {Verhaaren}}]{Burdman:2014zta}%
  \BibitemOpen
  \bibfield  {author} {\bibinfo {author} {\bibfnamefont {G.}~\bibnamefont
  {Burdman}}, \bibinfo {author} {\bibfnamefont {Z.}~\bibnamefont {Chacko}},
  \bibinfo {author} {\bibfnamefont {R.}~\bibnamefont {Harnik}}, \bibinfo
  {author} {\bibfnamefont {L.}~\bibnamefont {de~Lima}}, \ and\ \bibinfo
  {author} {\bibfnamefont {C.~B.}\ \bibnamefont {Verhaaren}},\ }\href {\doibase
  10.1103/PhysRevD.91.055007} {\bibfield  {journal} {\bibinfo  {journal} {Phys.
  Rev.}\ }\textbf {\bibinfo {volume} {D91}},\ \bibinfo {pages} {055007}
  (\bibinfo {year} {2015})},\ \Eprint {http://arxiv.org/abs/1411.3310}
  {arXiv:1411.3310 [hep-ph]} \BibitemShut {NoStop}%
%%CITATION = ARXIV:1411.3310;%%
\bibitem [{\citenamefont {ATLAS}(2020)}]{ATLAS:2020qdt}%
  \BibitemOpen
  \bibfield  {author} {\bibinfo {author} {\bibnamefont {ATLAS}},\ }\href@noop
  {} {\  (\bibinfo {year} {2020})},\ \Eprint
  {http://arxiv.org/abs/ATLAS-CONF-2020-027} {ATLAS-CONF-2020-027} \BibitemShut
  {NoStop}%
\bibitem [{\citenamefont {Sirunyan}\ \emph {et~al.}(2019)\citenamefont
  {Sirunyan} \emph {et~al.}}]{Sirunyan:2018koj}%
  \BibitemOpen
  \bibfield  {author} {\bibinfo {author} {\bibfnamefont {A.~M.}\ \bibnamefont
  {Sirunyan}} \emph {et~al.} (\bibinfo {collaboration} {CMS}),\ }\href
  {\doibase 10.1140/epjc/s10052-019-6909-y} {\bibfield  {journal} {\bibinfo
  {journal} {Eur. Phys. J. C}\ }\textbf {\bibinfo {volume} {79}},\ \bibinfo
  {pages} {421} (\bibinfo {year} {2019})},\ \Eprint
  {http://arxiv.org/abs/1809.10733} {arXiv:1809.10733 [hep-ex]} \BibitemShut
  {NoStop}%
\bibitem [{\citenamefont {Beauchesne}\ \emph {et~al.}(2016)\citenamefont
  {Beauchesne}, \citenamefont {Earl},\ and\ \citenamefont
  {Grégoire}}]{Beauchesne:2015lva}%
  \BibitemOpen
  \bibfield  {author} {\bibinfo {author} {\bibfnamefont {H.}~\bibnamefont
  {Beauchesne}}, \bibinfo {author} {\bibfnamefont {K.}~\bibnamefont {Earl}}, \
  and\ \bibinfo {author} {\bibfnamefont {T.}~\bibnamefont {Grégoire}},\ }\href
  {\doibase 10.1007/JHEP01(2016)130} {\bibfield  {journal} {\bibinfo  {journal}
  {JHEP}\ }\textbf {\bibinfo {volume} {01}},\ \bibinfo {pages} {130} (\bibinfo
  {year} {2016})},\ \Eprint {http://arxiv.org/abs/1510.06069} {arXiv:1510.06069
  [hep-ph]} \BibitemShut {NoStop}%
%%CITATION = ARXIV:1510.06069;%%
\bibitem [{\citenamefont {Harnik}\ \emph {et~al.}(2017)\citenamefont {Harnik},
  \citenamefont {Howe},\ and\ \citenamefont {Kearney}}]{Harnik:2016koz}%
  \BibitemOpen
  \bibfield  {author} {\bibinfo {author} {\bibfnamefont {R.}~\bibnamefont
  {Harnik}}, \bibinfo {author} {\bibfnamefont {K.}~\bibnamefont {Howe}}, \ and\
  \bibinfo {author} {\bibfnamefont {J.}~\bibnamefont {Kearney}},\ }\href
  {\doibase 10.1007/JHEP03(2017)111} {\bibfield  {journal} {\bibinfo  {journal}
  {JHEP}\ }\textbf {\bibinfo {volume} {03}},\ \bibinfo {pages} {111} (\bibinfo
  {year} {2017})},\ \Eprint {http://arxiv.org/abs/1603.03772} {arXiv:1603.03772
  [hep-ph]} \BibitemShut {NoStop}%
%%CITATION = ARXIV:1603.03772;%%
\bibitem [{\citenamefont {Cs\'aki}\ \emph {et~al.}(2020)\citenamefont
  {Cs\'aki}, \citenamefont {Guan}, \citenamefont {Ma},\ and\ \citenamefont
  {Shu}}]{Csaki:2019qgb}%
  \BibitemOpen
  \bibfield  {author} {\bibinfo {author} {\bibfnamefont {C.}~\bibnamefont
  {Cs\'aki}}, \bibinfo {author} {\bibfnamefont {C.-S.}\ \bibnamefont {Guan}},
  \bibinfo {author} {\bibfnamefont {T.}~\bibnamefont {Ma}}, \ and\ \bibinfo
  {author} {\bibfnamefont {J.}~\bibnamefont {Shu}},\ }\href {\doibase
  10.1007/JHEP12(2020)005} {\bibfield  {journal} {\bibinfo  {journal} {JHEP}\
  }\textbf {\bibinfo {volume} {12}},\ \bibinfo {pages} {005} (\bibinfo {year}
  {2020})},\ \Eprint {http://arxiv.org/abs/1910.14085} {arXiv:1910.14085
  [hep-ph]} \BibitemShut {NoStop}%
\bibitem [{\citenamefont {Dawson}\ \emph {et~al.}(2013)\citenamefont {Dawson}
  \emph {et~al.}}]{Dawson:2013bba}%
  \BibitemOpen
  \bibfield  {author} {\bibinfo {author} {\bibfnamefont {S.}~\bibnamefont
  {Dawson}} \emph {et~al.},\ }in\ \href@noop {} {\emph {\bibinfo {booktitle}
  {{Community Summer Study 2013}: {Snowmass on the Mississippi}}}}\ (\bibinfo
  {year} {2013})\ \Eprint {http://arxiv.org/abs/1310.8361} {arXiv:1310.8361
  [hep-ex]} \BibitemShut {NoStop}%
\bibitem [{\citenamefont {Craig}\ \emph {et~al.}(2013)\citenamefont {Craig},
  \citenamefont {Englert},\ and\ \citenamefont {McCullough}}]{Craig:2013xia}%
  \BibitemOpen
  \bibfield  {author} {\bibinfo {author} {\bibfnamefont {N.}~\bibnamefont
  {Craig}}, \bibinfo {author} {\bibfnamefont {C.}~\bibnamefont {Englert}}, \
  and\ \bibinfo {author} {\bibfnamefont {M.}~\bibnamefont {McCullough}},\
  }\href {\doibase 10.1103/PhysRevLett.111.121803} {\bibfield  {journal}
  {\bibinfo  {journal} {Phys. Rev. Lett.}\ }\textbf {\bibinfo {volume} {111}},\
  \bibinfo {pages} {121803} (\bibinfo {year} {2013})},\ \Eprint
  {http://arxiv.org/abs/1305.5251} {arXiv:1305.5251 [hep-ph]} \BibitemShut
  {NoStop}%
\bibitem [{\citenamefont {Curtin}\ and\ \citenamefont
  {Verhaaren}(2015)}]{Curtin:2015fna}%
  \BibitemOpen
  \bibfield  {author} {\bibinfo {author} {\bibfnamefont {D.}~\bibnamefont
  {Curtin}}\ and\ \bibinfo {author} {\bibfnamefont {C.~B.}\ \bibnamefont
  {Verhaaren}},\ }\href {\doibase 10.1007/JHEP12(2015)072} {\bibfield
  {journal} {\bibinfo  {journal} {JHEP}\ }\textbf {\bibinfo {volume} {12}},\
  \bibinfo {pages} {072} (\bibinfo {year} {2015})},\ \Eprint
  {http://arxiv.org/abs/1506.06141} {arXiv:1506.06141 [hep-ph]} \BibitemShut
  {NoStop}%
%%CITATION = ARXIV:1506.06141;%%
\bibitem [{\citenamefont {Aghanim}\ \emph {et~al.}(2018)\citenamefont {Aghanim}
  \emph {et~al.}}]{Aghanim:2018eyx}%
  \BibitemOpen
  \bibfield  {author} {\bibinfo {author} {\bibfnamefont {N.}~\bibnamefont
  {Aghanim}} \emph {et~al.} (\bibinfo {collaboration} {Planck}),\ }\href@noop
  {} {\  (\bibinfo {year} {2018})},\ \Eprint {http://arxiv.org/abs/1807.06209}
  {arXiv:1807.06209 [astro-ph.CO]} \BibitemShut {NoStop}%
%%CITATION = ARXIV:1807.06209;%%
\bibitem [{\citenamefont {Garcia~Garcia}\ \emph
  {et~al.}(2015{\natexlab{a}})\citenamefont {Garcia~Garcia}, \citenamefont
  {Lasenby},\ and\ \citenamefont {March-Russell}}]{Garcia:2015loa}%
  \BibitemOpen
  \bibfield  {author} {\bibinfo {author} {\bibfnamefont {I.}~\bibnamefont
  {Garcia~Garcia}}, \bibinfo {author} {\bibfnamefont {R.}~\bibnamefont
  {Lasenby}}, \ and\ \bibinfo {author} {\bibfnamefont {J.}~\bibnamefont
  {March-Russell}},\ }\href {\doibase 10.1103/PhysRevD.92.055034} {\bibfield
  {journal} {\bibinfo  {journal} {Phys. Rev.}\ }\textbf {\bibinfo {volume}
  {D92}},\ \bibinfo {pages} {055034} (\bibinfo {year} {2015}{\natexlab{a}})},\
  \Eprint {http://arxiv.org/abs/1505.07109} {arXiv:1505.07109 [hep-ph]}
  \BibitemShut {NoStop}%
%%CITATION = ARXIV:1505.07109;%%
\bibitem [{\citenamefont {Chacko}\ \emph {et~al.}(2017)\citenamefont {Chacko},
  \citenamefont {Craig}, \citenamefont {Fox},\ and\ \citenamefont
  {Harnik}}]{Chacko:2016hvu}%
  \BibitemOpen
  \bibfield  {author} {\bibinfo {author} {\bibfnamefont {Z.}~\bibnamefont
  {Chacko}}, \bibinfo {author} {\bibfnamefont {N.}~\bibnamefont {Craig}},
  \bibinfo {author} {\bibfnamefont {P.~J.}\ \bibnamefont {Fox}}, \ and\
  \bibinfo {author} {\bibfnamefont {R.}~\bibnamefont {Harnik}},\ }\href
  {\doibase 10.1007/JHEP07(2017)023} {\bibfield  {journal} {\bibinfo  {journal}
  {JHEP}\ }\textbf {\bibinfo {volume} {07}},\ \bibinfo {pages} {023} (\bibinfo
  {year} {2017})},\ \Eprint {http://arxiv.org/abs/1611.07975} {arXiv:1611.07975
  [hep-ph]} \BibitemShut {NoStop}%
%%CITATION = ARXIV:1611.07975;%%
\bibitem [{\citenamefont {Craig}\ \emph {et~al.}(2017)\citenamefont {Craig},
  \citenamefont {Koren},\ and\ \citenamefont {Trott}}]{Craig:2016lyx}%
  \BibitemOpen
  \bibfield  {author} {\bibinfo {author} {\bibfnamefont {N.}~\bibnamefont
  {Craig}}, \bibinfo {author} {\bibfnamefont {S.}~\bibnamefont {Koren}}, \ and\
  \bibinfo {author} {\bibfnamefont {T.}~\bibnamefont {Trott}},\ }\href
  {\doibase 10.1007/JHEP05(2017)038} {\bibfield  {journal} {\bibinfo  {journal}
  {JHEP}\ }\textbf {\bibinfo {volume} {05}},\ \bibinfo {pages} {038} (\bibinfo
  {year} {2017})},\ \Eprint {http://arxiv.org/abs/1611.07977} {arXiv:1611.07977
  [hep-ph]} \BibitemShut {NoStop}%
%%CITATION = ARXIV:1611.07977;%%
\bibitem [{\citenamefont {Craig}\ and\ \citenamefont
  {Katz}(2015)}]{Craig:2015xla}%
  \BibitemOpen
  \bibfield  {author} {\bibinfo {author} {\bibfnamefont {N.}~\bibnamefont
  {Craig}}\ and\ \bibinfo {author} {\bibfnamefont {A.}~\bibnamefont {Katz}},\
  }\href {\doibase 10.1088/1475-7516/2015/10/054} {\bibfield  {journal}
  {\bibinfo  {journal} {JCAP}\ }\textbf {\bibinfo {volume} {1510}},\ \bibinfo
  {pages} {054} (\bibinfo {year} {2015})},\ \Eprint
  {http://arxiv.org/abs/1505.07113} {arXiv:1505.07113 [hep-ph]} \BibitemShut
  {NoStop}%
%%CITATION = ARXIV:1505.07113;%%
\bibitem [{\citenamefont {Farina}(2015)}]{Farina:2015uea}%
  \BibitemOpen
  \bibfield  {author} {\bibinfo {author} {\bibfnamefont {M.}~\bibnamefont
  {Farina}},\ }\href {\doibase 10.1088/1475-7516/2015/11/017} {\bibfield
  {journal} {\bibinfo  {journal} {JCAP}\ }\textbf {\bibinfo {volume} {1511}},\
  \bibinfo {pages} {017} (\bibinfo {year} {2015})},\ \Eprint
  {http://arxiv.org/abs/1506.03520} {arXiv:1506.03520 [hep-ph]} \BibitemShut
  {NoStop}%
%%CITATION = ARXIV:1506.03520;%%
\bibitem [{\citenamefont {Garcia~Garcia}\ \emph
  {et~al.}(2015{\natexlab{b}})\citenamefont {Garcia~Garcia}, \citenamefont
  {Lasenby},\ and\ \citenamefont {March-Russell}}]{Garcia:2015toa}%
  \BibitemOpen
  \bibfield  {author} {\bibinfo {author} {\bibfnamefont {I.}~\bibnamefont
  {Garcia~Garcia}}, \bibinfo {author} {\bibfnamefont {R.}~\bibnamefont
  {Lasenby}}, \ and\ \bibinfo {author} {\bibfnamefont {J.}~\bibnamefont
  {March-Russell}},\ }\href {\doibase 10.1103/PhysRevLett.115.121801}
  {\bibfield  {journal} {\bibinfo  {journal} {Phys. Rev. Lett.}\ }\textbf
  {\bibinfo {volume} {115}},\ \bibinfo {pages} {121801} (\bibinfo {year}
  {2015}{\natexlab{b}})},\ \Eprint {http://arxiv.org/abs/1505.07410}
  {arXiv:1505.07410 [hep-ph]} \BibitemShut {NoStop}%
%%CITATION = ARXIV:1505.07410;%%
\bibitem [{\citenamefont {Farina}\ \emph {et~al.}(2016)\citenamefont {Farina},
  \citenamefont {Monteux},\ and\ \citenamefont {Shin}}]{Farina:2016ndq}%
  \BibitemOpen
  \bibfield  {author} {\bibinfo {author} {\bibfnamefont {M.}~\bibnamefont
  {Farina}}, \bibinfo {author} {\bibfnamefont {A.}~\bibnamefont {Monteux}}, \
  and\ \bibinfo {author} {\bibfnamefont {C.~S.}\ \bibnamefont {Shin}},\ }\href
  {\doibase 10.1103/PhysRevD.94.035017} {\bibfield  {journal} {\bibinfo
  {journal} {Phys. Rev.}\ }\textbf {\bibinfo {volume} {D94}},\ \bibinfo {pages}
  {035017} (\bibinfo {year} {2016})},\ \Eprint
  {http://arxiv.org/abs/1604.08211} {arXiv:1604.08211 [hep-ph]} \BibitemShut
  {NoStop}%
%%CITATION = ARXIV:1604.08211;%%
\bibitem [{\citenamefont {Freytsis}\ \emph {et~al.}(2016)\citenamefont
  {Freytsis}, \citenamefont {Knapen}, \citenamefont {Robinson},\ and\
  \citenamefont {Tsai}}]{Freytsis:2016dgf}%
  \BibitemOpen
  \bibfield  {author} {\bibinfo {author} {\bibfnamefont {M.}~\bibnamefont
  {Freytsis}}, \bibinfo {author} {\bibfnamefont {S.}~\bibnamefont {Knapen}},
  \bibinfo {author} {\bibfnamefont {D.~J.}\ \bibnamefont {Robinson}}, \ and\
  \bibinfo {author} {\bibfnamefont {Y.}~\bibnamefont {Tsai}},\ }\href {\doibase
  10.1007/JHEP05(2016)018} {\bibfield  {journal} {\bibinfo  {journal} {JHEP}\
  }\textbf {\bibinfo {volume} {05}},\ \bibinfo {pages} {018} (\bibinfo {year}
  {2016})},\ \Eprint {http://arxiv.org/abs/1601.07556} {arXiv:1601.07556
  [hep-ph]} \BibitemShut {NoStop}%
\bibitem [{\citenamefont {Cheng}\ \emph
  {et~al.}(2018{\natexlab{b}})\citenamefont {Cheng}, \citenamefont {Li},\ and\
  \citenamefont {Zheng}}]{Cheng:2018vaj}%
  \BibitemOpen
  \bibfield  {author} {\bibinfo {author} {\bibfnamefont {H.-C.}\ \bibnamefont
  {Cheng}}, \bibinfo {author} {\bibfnamefont {L.}~\bibnamefont {Li}}, \ and\
  \bibinfo {author} {\bibfnamefont {R.}~\bibnamefont {Zheng}},\ }\href
  {\doibase 10.1007/JHEP09(2018)098} {\bibfield  {journal} {\bibinfo  {journal}
  {JHEP}\ }\textbf {\bibinfo {volume} {09}},\ \bibinfo {pages} {098} (\bibinfo
  {year} {2018}{\natexlab{b}})},\ \Eprint {http://arxiv.org/abs/1805.12139}
  {arXiv:1805.12139 [hep-ph]} \BibitemShut {NoStop}%
%%CITATION = ARXIV:1805.12139;%%
\bibitem [{\citenamefont {Hochberg}\ \emph {et~al.}(2019)\citenamefont
  {Hochberg}, \citenamefont {Kuflik},\ and\ \citenamefont
  {Murayama}}]{Hochberg:2018vdo}%
  \BibitemOpen
  \bibfield  {author} {\bibinfo {author} {\bibfnamefont {Y.}~\bibnamefont
  {Hochberg}}, \bibinfo {author} {\bibfnamefont {E.}~\bibnamefont {Kuflik}}, \
  and\ \bibinfo {author} {\bibfnamefont {H.}~\bibnamefont {Murayama}},\ }\href
  {\doibase 10.1103/PhysRevD.99.015005} {\bibfield  {journal} {\bibinfo
  {journal} {Phys. Rev.}\ }\textbf {\bibinfo {volume} {D99}},\ \bibinfo {pages}
  {015005} (\bibinfo {year} {2019})},\ \Eprint
  {http://arxiv.org/abs/1805.09345} {arXiv:1805.09345 [hep-ph]} \BibitemShut
  {NoStop}%
%%CITATION = ARXIV:1805.09345;%%
\bibitem [{\citenamefont {Badziak}\ \emph {et~al.}(2020)\citenamefont
  {Badziak}, \citenamefont {Grilli Di~Cortona},\ and\ \citenamefont
  {Harigaya}}]{Badziak:2019zys}%
  \BibitemOpen
  \bibfield  {author} {\bibinfo {author} {\bibfnamefont {M.}~\bibnamefont
  {Badziak}}, \bibinfo {author} {\bibfnamefont {G.}~\bibnamefont {Grilli
  Di~Cortona}}, \ and\ \bibinfo {author} {\bibfnamefont {K.}~\bibnamefont
  {Harigaya}},\ }\href {\doibase 10.1103/PhysRevLett.124.121803} {\bibfield
  {journal} {\bibinfo  {journal} {Phys. Rev. Lett.}\ }\textbf {\bibinfo
  {volume} {124}},\ \bibinfo {pages} {121803} (\bibinfo {year} {2020})},\
  \Eprint {http://arxiv.org/abs/1911.03481} {arXiv:1911.03481 [hep-ph]}
  \BibitemShut {NoStop}%
\bibitem [{\citenamefont {Koren}\ and\ \citenamefont
  {McGehee}(2019)}]{Koren:2019iuv}%
  \BibitemOpen
  \bibfield  {author} {\bibinfo {author} {\bibfnamefont {S.}~\bibnamefont
  {Koren}}\ and\ \bibinfo {author} {\bibfnamefont {R.}~\bibnamefont
  {McGehee}},\ }\href@noop {} {\  (\bibinfo {year} {2019})},\ \Eprint
  {http://arxiv.org/abs/1908.03559} {arXiv:1908.03559 [hep-ph]} \BibitemShut
  {NoStop}%
%%CITATION = ARXIV:1908.03559;%%
\bibitem [{\citenamefont {Terning}\ \emph {et~al.}(2019)\citenamefont
  {Terning}, \citenamefont {Verhaaren},\ and\ \citenamefont
  {Zora}}]{Terning:2019hgj}%
  \BibitemOpen
  \bibfield  {author} {\bibinfo {author} {\bibfnamefont {J.}~\bibnamefont
  {Terning}}, \bibinfo {author} {\bibfnamefont {C.~B.}\ \bibnamefont
  {Verhaaren}}, \ and\ \bibinfo {author} {\bibfnamefont {K.}~\bibnamefont
  {Zora}},\ }\href {\doibase 10.1103/PhysRevD.99.095020} {\bibfield  {journal}
  {\bibinfo  {journal} {Phys. Rev. D}\ }\textbf {\bibinfo {volume} {99}},\
  \bibinfo {pages} {095020} (\bibinfo {year} {2019})},\ \Eprint
  {http://arxiv.org/abs/1902.08211} {arXiv:1902.08211 [hep-ph]} \BibitemShut
  {NoStop}%
\bibitem [{\citenamefont {Beauchesne}(2020)}]{Beauchesne:2020mih}%
  \BibitemOpen
  \bibfield  {author} {\bibinfo {author} {\bibfnamefont {H.}~\bibnamefont
  {Beauchesne}},\ }\href {\doibase 10.1007/JHEP09(2020)048} {\bibfield
  {journal} {\bibinfo  {journal} {JHEP}\ }\textbf {\bibinfo {volume} {09}},\
  \bibinfo {pages} {048} (\bibinfo {year} {2020})},\ \Eprint
  {http://arxiv.org/abs/2007.00052} {arXiv:2007.00052 [hep-ph]} \BibitemShut
  {NoStop}%
\bibitem [{\citenamefont {Ahmed}\ \emph {et~al.}(2020)\citenamefont {Ahmed},
  \citenamefont {Najjari},\ and\ \citenamefont {Verhaaren}}]{Ahmed:2020hiw}%
  \BibitemOpen
  \bibfield  {author} {\bibinfo {author} {\bibfnamefont {A.}~\bibnamefont
  {Ahmed}}, \bibinfo {author} {\bibfnamefont {S.}~\bibnamefont {Najjari}}, \
  and\ \bibinfo {author} {\bibfnamefont {C.~B.}\ \bibnamefont {Verhaaren}},\
  }\href {\doibase 10.1007/JHEP06(2020)007} {\bibfield  {journal} {\bibinfo
  {journal} {JHEP}\ }\textbf {\bibinfo {volume} {06}},\ \bibinfo {pages} {007}
  (\bibinfo {year} {2020})},\ \Eprint {http://arxiv.org/abs/2003.08947}
  {arXiv:2003.08947 [hep-ph]} \BibitemShut {NoStop}%
\bibitem [{\citenamefont {Curtin}\ and\ \citenamefont
  {Gryba}(2021)}]{Curtin:2021alk}%
  \BibitemOpen
  \bibfield  {author} {\bibinfo {author} {\bibfnamefont {D.}~\bibnamefont
  {Curtin}}\ and\ \bibinfo {author} {\bibfnamefont {S.}~\bibnamefont {Gryba}},\
  }\href@noop {} {\  (\bibinfo {year} {2021})},\ \Eprint
  {http://arxiv.org/abs/2101.11019} {arXiv:2101.11019 [hep-ph]} \BibitemShut
  {NoStop}%
\bibitem [{\citenamefont {Ritter}\ and\ \citenamefont
  {Volkas}(2021)}]{Ritter:2021hgu}%
  \BibitemOpen
  \bibfield  {author} {\bibinfo {author} {\bibfnamefont {A.~C.}\ \bibnamefont
  {Ritter}}\ and\ \bibinfo {author} {\bibfnamefont {R.~R.}\ \bibnamefont
  {Volkas}},\ }\href@noop {} {\  (\bibinfo {year} {2021})},\ \Eprint
  {http://arxiv.org/abs/2101.07421} {arXiv:2101.07421 [hep-ph]} \BibitemShut
  {NoStop}%
\bibitem [{\citenamefont {Batell}\ and\ \citenamefont
  {Verhaaren}(2019)}]{Batell:2019ptb}%
  \BibitemOpen
  \bibfield  {author} {\bibinfo {author} {\bibfnamefont {B.}~\bibnamefont
  {Batell}}\ and\ \bibinfo {author} {\bibfnamefont {C.~B.}\ \bibnamefont
  {Verhaaren}},\ }\href@noop {} {\  (\bibinfo {year} {2019})},\ \Eprint
  {http://arxiv.org/abs/1904.10468} {arXiv:1904.10468 [hep-ph]} \BibitemShut
  {NoStop}%
%%CITATION = ARXIV:1904.10468;%%
\bibitem [{CMS(2017)}]{CMS:2017pet}%
  \BibitemOpen
  \href@noop {} {\  (\bibinfo {year} {2017})}\BibitemShut {NoStop}%
\bibitem [{\citenamefont {Craig}\ \emph
  {et~al.}(2015{\natexlab{b}})\citenamefont {Craig}, \citenamefont {Knapen},\
  and\ \citenamefont {Longhi}}]{Craig:2014roa}%
  \BibitemOpen
  \bibfield  {author} {\bibinfo {author} {\bibfnamefont {N.}~\bibnamefont
  {Craig}}, \bibinfo {author} {\bibfnamefont {S.}~\bibnamefont {Knapen}}, \
  and\ \bibinfo {author} {\bibfnamefont {P.}~\bibnamefont {Longhi}},\ }\href
  {\doibase 10.1007/JHEP03(2015)106} {\bibfield  {journal} {\bibinfo  {journal}
  {JHEP}\ }\textbf {\bibinfo {volume} {03}},\ \bibinfo {pages} {106} (\bibinfo
  {year} {2015}{\natexlab{b}})},\ \Eprint {http://arxiv.org/abs/1411.7393}
  {arXiv:1411.7393 [hep-ph]} \BibitemShut {NoStop}%
\bibitem [{Note1()}]{Note1}%
  \BibitemOpen
  \bibinfo {note} {Even though we assume twin bottom and glueball masses
  expected from the $\protect \mathbb {Z}_2$ symmetry, their exact masses are
  not important for the viability of this scenario, as long as they can
  annihilate or decay into the visible sector.}\BibitemShut {Stop}%
\bibitem [{\citenamefont {Goodsell}\ and\ \citenamefont
  {Staub}(2018)}]{Goodsell:2018tti}%
  \BibitemOpen
  \bibfield  {author} {\bibinfo {author} {\bibfnamefont {M.~D.}\ \bibnamefont
  {Goodsell}}\ and\ \bibinfo {author} {\bibfnamefont {F.}~\bibnamefont
  {Staub}},\ }\href {\doibase 10.1140/epjc/s10052-018-6127-z} {\bibfield
  {journal} {\bibinfo  {journal} {Eur. Phys. J. C}\ }\textbf {\bibinfo {volume}
  {78}},\ \bibinfo {pages} {649} (\bibinfo {year} {2018})},\ \Eprint
  {http://arxiv.org/abs/1805.07306} {arXiv:1805.07306 [hep-ph]} \BibitemShut
  {NoStop}%
\bibitem [{Note2()}]{Note2}%
  \BibitemOpen
  \bibinfo {note} {Note that the Majorana fermion nature of the $\tau '_1$
  introduces an extra factor of 2 into the Feynman rule associated with this
  vertex.}\BibitemShut {Stop}%
\bibitem [{\citenamefont {Akerib}\ \emph {et~al.}(2015)\citenamefont {Akerib}
  \emph {et~al.}}]{Akerib:2015cja}%
  \BibitemOpen
  \bibfield  {author} {\bibinfo {author} {\bibfnamefont {D.~S.}\ \bibnamefont
  {Akerib}} \emph {et~al.} (\bibinfo {collaboration} {LZ}),\ }\href@noop {} {\
  (\bibinfo {year} {2015})},\ \Eprint {http://arxiv.org/abs/1509.02910}
  {arXiv:1509.02910 [physics.ins-det]} \BibitemShut {NoStop}%
%%CITATION = ARXIV:1509.02910;%%
\bibitem [{\citenamefont {Aprile}\ \emph {et~al.}()\citenamefont {Aprile} \emph
  {et~al.}}]{Aprile:2015uzo}%
  \BibitemOpen
  \bibfield  {author} {\bibinfo {author} {\bibfnamefont {E.}~\bibnamefont
  {Aprile}} \emph {et~al.} (\bibinfo {collaboration} {XENON}),\ }\href@noop {}
  {\ }\BibitemShut {NoStop}%
\bibitem [{\citenamefont {Aalbers}\ \emph {et~al.}(2016)\citenamefont {Aalbers}
  \emph {et~al.}}]{Aalbers:2016jon}%
  \BibitemOpen
  \bibfield  {author} {\bibinfo {author} {\bibfnamefont {J.}~\bibnamefont
  {Aalbers}} \emph {et~al.} (\bibinfo {collaboration} {DARWIN}),\ }\href
  {\doibase 10.1088/1475-7516/2016/11/017} {\bibfield  {journal} {\bibinfo
  {journal} {JCAP}\ }\textbf {\bibinfo {volume} {11}},\ \bibinfo {pages} {017}
  (\bibinfo {year} {2016})},\ \Eprint {http://arxiv.org/abs/1606.07001}
  {arXiv:1606.07001 [astro-ph.IM]} \BibitemShut {NoStop}%
\bibitem [{\citenamefont {Okun}(1979)}]{Okun:1980kw}%
  \BibitemOpen
  \bibfield  {author} {\bibinfo {author} {\bibfnamefont {L.~B.}\ \bibnamefont
  {Okun}},\ }\href@noop {} {\bibfield  {journal} {\bibinfo  {journal} {Pisma
  Zh. Eksp. Teor. Fiz.}\ }\textbf {\bibinfo {volume} {31}},\ \bibinfo {pages}
  {156} (\bibinfo {year} {1979})}\BibitemShut {NoStop}%
\bibitem [{\citenamefont {Okun}(1980)}]{Okun:1980mu}%
  \BibitemOpen
  \bibfield  {author} {\bibinfo {author} {\bibfnamefont {L.~B.}\ \bibnamefont
  {Okun}},\ }\href {\doibase 10.1016/0550-3213(80)90439-3} {\bibfield
  {journal} {\bibinfo  {journal} {Nucl. Phys. B}\ }\textbf {\bibinfo {volume}
  {173}},\ \bibinfo {pages} {1} (\bibinfo {year} {1980})}\BibitemShut {NoStop}%
\bibitem [{\citenamefont {Kang}\ and\ \citenamefont
  {Luty}(2009)}]{Kang:2008ea}%
  \BibitemOpen
  \bibfield  {author} {\bibinfo {author} {\bibfnamefont {J.}~\bibnamefont
  {Kang}}\ and\ \bibinfo {author} {\bibfnamefont {M.~A.}\ \bibnamefont
  {Luty}},\ }\href {\doibase 10.1088/1126-6708/2009/11/065} {\bibfield
  {journal} {\bibinfo  {journal} {JHEP}\ }\textbf {\bibinfo {volume} {11}},\
  \bibinfo {pages} {065} (\bibinfo {year} {2009})},\ \Eprint
  {http://arxiv.org/abs/0805.4642} {arXiv:0805.4642 [hep-ph]} \BibitemShut
  {NoStop}%
\bibitem [{\citenamefont {Daylan}\ \emph {et~al.}(2016)\citenamefont {Daylan},
  \citenamefont {Finkbeiner}, \citenamefont {Hooper}, \citenamefont {Linden},
  \citenamefont {Portillo}, \citenamefont {Rodd},\ and\ \citenamefont
  {Slatyer}}]{Daylan:2014rsa}%
  \BibitemOpen
  \bibfield  {author} {\bibinfo {author} {\bibfnamefont {T.}~\bibnamefont
  {Daylan}}, \bibinfo {author} {\bibfnamefont {D.~P.}\ \bibnamefont
  {Finkbeiner}}, \bibinfo {author} {\bibfnamefont {D.}~\bibnamefont {Hooper}},
  \bibinfo {author} {\bibfnamefont {T.}~\bibnamefont {Linden}}, \bibinfo
  {author} {\bibfnamefont {S.~K.~N.}\ \bibnamefont {Portillo}}, \bibinfo
  {author} {\bibfnamefont {N.~L.}\ \bibnamefont {Rodd}}, \ and\ \bibinfo
  {author} {\bibfnamefont {T.~R.}\ \bibnamefont {Slatyer}},\ }\href {\doibase
  10.1016/j.dark.2015.12.005} {\bibfield  {journal} {\bibinfo  {journal} {Phys.
  Dark Univ.}\ }\textbf {\bibinfo {volume} {12}},\ \bibinfo {pages} {1}
  (\bibinfo {year} {2016})},\ \Eprint {http://arxiv.org/abs/1402.6703}
  {arXiv:1402.6703 [astro-ph.HE]} \BibitemShut {NoStop}%
\bibitem [{\citenamefont {Hooper}\ and\ \citenamefont
  {Goodenough}(2011)}]{Hooper:2010mq}%
  \BibitemOpen
  \bibfield  {author} {\bibinfo {author} {\bibfnamefont {D.}~\bibnamefont
  {Hooper}}\ and\ \bibinfo {author} {\bibfnamefont {L.}~\bibnamefont
  {Goodenough}},\ }\href {\doibase 10.1016/j.physletb.2011.02.029} {\bibfield
  {journal} {\bibinfo  {journal} {Phys. Lett. B}\ }\textbf {\bibinfo {volume}
  {697}},\ \bibinfo {pages} {412} (\bibinfo {year} {2011})},\ \Eprint
  {http://arxiv.org/abs/1010.2752} {arXiv:1010.2752 [hep-ph]} \BibitemShut
  {NoStop}%
\bibitem [{\citenamefont {Cuoco}\ \emph {et~al.}(2017)\citenamefont {Cuoco},
  \citenamefont {Kr\"amer},\ and\ \citenamefont {Korsmeier}}]{Cuoco:2016eej}%
  \BibitemOpen
  \bibfield  {author} {\bibinfo {author} {\bibfnamefont {A.}~\bibnamefont
  {Cuoco}}, \bibinfo {author} {\bibfnamefont {M.}~\bibnamefont {Kr\"amer}}, \
  and\ \bibinfo {author} {\bibfnamefont {M.}~\bibnamefont {Korsmeier}},\ }\href
  {\doibase 10.1103/PhysRevLett.118.191102} {\bibfield  {journal} {\bibinfo
  {journal} {Phys. Rev. Lett.}\ }\textbf {\bibinfo {volume} {118}},\ \bibinfo
  {pages} {191102} (\bibinfo {year} {2017})},\ \Eprint
  {http://arxiv.org/abs/1610.03071} {arXiv:1610.03071 [astro-ph.HE]}
  \BibitemShut {NoStop}%
\bibitem [{\citenamefont {Cui}\ \emph {et~al.}(2017{\natexlab{b}})\citenamefont
  {Cui}, \citenamefont {Yuan}, \citenamefont {Tsai},\ and\ \citenamefont
  {Fan}}]{Cui:2016ppb}%
  \BibitemOpen
  \bibfield  {author} {\bibinfo {author} {\bibfnamefont {M.-Y.}\ \bibnamefont
  {Cui}}, \bibinfo {author} {\bibfnamefont {Q.}~\bibnamefont {Yuan}}, \bibinfo
  {author} {\bibfnamefont {Y.-L.~S.}\ \bibnamefont {Tsai}}, \ and\ \bibinfo
  {author} {\bibfnamefont {Y.-Z.}\ \bibnamefont {Fan}},\ }\href {\doibase
  10.1103/PhysRevLett.118.191101} {\bibfield  {journal} {\bibinfo  {journal}
  {Phys. Rev. Lett.}\ }\textbf {\bibinfo {volume} {118}},\ \bibinfo {pages}
  {191101} (\bibinfo {year} {2017}{\natexlab{b}})},\ \Eprint
  {http://arxiv.org/abs/1610.03840} {arXiv:1610.03840 [astro-ph.HE]}
  \BibitemShut {NoStop}%
\bibitem [{\citenamefont {Cholis}\ \emph {et~al.}(2019)\citenamefont {Cholis},
  \citenamefont {Linden},\ and\ \citenamefont {Hooper}}]{Cholis:2019ejx}%
  \BibitemOpen
  \bibfield  {author} {\bibinfo {author} {\bibfnamefont {I.}~\bibnamefont
  {Cholis}}, \bibinfo {author} {\bibfnamefont {T.}~\bibnamefont {Linden}}, \
  and\ \bibinfo {author} {\bibfnamefont {D.}~\bibnamefont {Hooper}},\ }\href
  {\doibase 10.1103/PhysRevD.99.103026} {\bibfield  {journal} {\bibinfo
  {journal} {Phys. Rev. D}\ }\textbf {\bibinfo {volume} {99}},\ \bibinfo
  {pages} {103026} (\bibinfo {year} {2019})},\ \Eprint
  {http://arxiv.org/abs/1903.02549} {arXiv:1903.02549 [astro-ph.HE]}
  \BibitemShut {NoStop}%
\bibitem [{\citenamefont {Curtin}\ and\ \citenamefont {Gemmell}(2021)}]{caleb}%
  \BibitemOpen
  \bibfield  {author} {\bibinfo {author} {\bibfnamefont {D.}~\bibnamefont
  {Curtin}}\ and\ \bibinfo {author} {\bibfnamefont {C.}~\bibnamefont
  {Gemmell}},\ }\href@noop {} {\bibfield  {journal} {\bibinfo  {journal} {In
  preparation.}\ } (\bibinfo {year} {2021})}\BibitemShut {NoStop}%
\bibitem [{\citenamefont {Abazajian}\ \emph {et~al.}(2016)\citenamefont
  {Abazajian} \emph {et~al.}}]{Abazajian:2016yjj}%
  \BibitemOpen
  \bibfield  {author} {\bibinfo {author} {\bibfnamefont {K.~N.}\ \bibnamefont
  {Abazajian}} \emph {et~al.} (\bibinfo {collaboration} {CMB-S4}),\ }\href@noop
  {} {\  (\bibinfo {year} {2016})},\ \Eprint {http://arxiv.org/abs/1610.02743}
  {arXiv:1610.02743 [astro-ph.CO]} \BibitemShut {NoStop}%
\bibitem [{\citenamefont {Bennett}\ \emph {et~al.}(2006)\citenamefont {Bennett}
  \emph {et~al.}}]{Bennett:2006fi}%
  \BibitemOpen
  \bibfield  {author} {\bibinfo {author} {\bibfnamefont {G.~W.}\ \bibnamefont
  {Bennett}} \emph {et~al.} (\bibinfo {collaboration} {Muon g-2}),\ }\href
  {\doibase 10.1103/PhysRevD.73.072003} {\bibfield  {journal} {\bibinfo
  {journal} {Phys. Rev. D}\ }\textbf {\bibinfo {volume} {73}},\ \bibinfo
  {pages} {072003} (\bibinfo {year} {2006})},\ \Eprint
  {http://arxiv.org/abs/hep-ex/0602035} {arXiv:hep-ex/0602035} \BibitemShut
  {NoStop}%
\bibitem [{\citenamefont {Abi}\ \emph {et~al.}(2021)\citenamefont {Abi} \emph
  {et~al.}}]{Abi:2021gix}%
  \BibitemOpen
  \bibfield  {author} {\bibinfo {author} {\bibfnamefont {B.}~\bibnamefont
  {Abi}} \emph {et~al.} (\bibinfo {collaboration} {Muon g-2}),\ }\href
  {\doibase 10.1103/PhysRevLett.126.141801} {\bibfield  {journal} {\bibinfo
  {journal} {Phys. Rev. Lett.}\ }\textbf {\bibinfo {volume} {126}},\ \bibinfo
  {pages} {141801} (\bibinfo {year} {2021})},\ \Eprint
  {http://arxiv.org/abs/2104.03281} {arXiv:2104.03281 [hep-ex]} \BibitemShut
  {NoStop}%
\bibitem [{\citenamefont {Fuks}\ \emph {et~al.}(2014)\citenamefont {Fuks},
  \citenamefont {Klasen}, \citenamefont {Lamprea},\ and\ \citenamefont
  {Rothering}}]{Fuks:2013lya}%
  \BibitemOpen
  \bibfield  {author} {\bibinfo {author} {\bibfnamefont {B.}~\bibnamefont
  {Fuks}}, \bibinfo {author} {\bibfnamefont {M.}~\bibnamefont {Klasen}},
  \bibinfo {author} {\bibfnamefont {D.~R.}\ \bibnamefont {Lamprea}}, \ and\
  \bibinfo {author} {\bibfnamefont {M.}~\bibnamefont {Rothering}},\ }\href
  {\doibase 10.1007/JHEP01(2014)168} {\bibfield  {journal} {\bibinfo  {journal}
  {JHEP}\ }\textbf {\bibinfo {volume} {01}},\ \bibinfo {pages} {168} (\bibinfo
  {year} {2014})},\ \Eprint {http://arxiv.org/abs/1310.2621} {arXiv:1310.2621
  [hep-ph]} \BibitemShut {NoStop}%
\bibitem [{\citenamefont {Chacko}\ \emph {et~al.}(2019)\citenamefont {Chacko},
  \citenamefont {Kilic}, \citenamefont {Najjari},\ and\ \citenamefont
  {Verhaaren}}]{Chacko:2019jgi}%
  \BibitemOpen
  \bibfield  {author} {\bibinfo {author} {\bibfnamefont {Z.}~\bibnamefont
  {Chacko}}, \bibinfo {author} {\bibfnamefont {C.}~\bibnamefont {Kilic}},
  \bibinfo {author} {\bibfnamefont {S.}~\bibnamefont {Najjari}}, \ and\
  \bibinfo {author} {\bibfnamefont {C.~B.}\ \bibnamefont {Verhaaren}},\ }\href
  {\doibase 10.1103/PhysRevD.100.035037} {\bibfield  {journal} {\bibinfo
  {journal} {Phys. Rev. D}\ }\textbf {\bibinfo {volume} {100}},\ \bibinfo
  {pages} {035037} (\bibinfo {year} {2019})},\ \Eprint
  {http://arxiv.org/abs/1904.11990} {arXiv:1904.11990 [hep-ph]} \BibitemShut
  {NoStop}%
\bibitem [{\citenamefont {Csaki}\ \emph {et~al.}(2015)\citenamefont {Csaki},
  \citenamefont {Kuflik}, \citenamefont {Lombardo},\ and\ \citenamefont
  {Slone}}]{Csaki:2015fba}%
  \BibitemOpen
  \bibfield  {author} {\bibinfo {author} {\bibfnamefont {C.}~\bibnamefont
  {Csaki}}, \bibinfo {author} {\bibfnamefont {E.}~\bibnamefont {Kuflik}},
  \bibinfo {author} {\bibfnamefont {S.}~\bibnamefont {Lombardo}}, \ and\
  \bibinfo {author} {\bibfnamefont {O.}~\bibnamefont {Slone}},\ }\href
  {\doibase 10.1103/PhysRevD.92.073008} {\bibfield  {journal} {\bibinfo
  {journal} {Phys. Rev. D}\ }\textbf {\bibinfo {volume} {92}},\ \bibinfo
  {pages} {073008} (\bibinfo {year} {2015})},\ \Eprint
  {http://arxiv.org/abs/1508.01522} {arXiv:1508.01522 [hep-ph]} \BibitemShut
  {NoStop}%
\bibitem [{\citenamefont {Cheng}\ \emph {et~al.}(2016)\citenamefont {Cheng},
  \citenamefont {Jung}, \citenamefont {Salvioni},\ and\ \citenamefont
  {Tsai}}]{Cheng:2015buv}%
  \BibitemOpen
  \bibfield  {author} {\bibinfo {author} {\bibfnamefont {H.-C.}\ \bibnamefont
  {Cheng}}, \bibinfo {author} {\bibfnamefont {S.}~\bibnamefont {Jung}},
  \bibinfo {author} {\bibfnamefont {E.}~\bibnamefont {Salvioni}}, \ and\
  \bibinfo {author} {\bibfnamefont {Y.}~\bibnamefont {Tsai}},\ }\href {\doibase
  10.1007/JHEP03(2016)074} {\bibfield  {journal} {\bibinfo  {journal} {JHEP}\
  }\textbf {\bibinfo {volume} {03}},\ \bibinfo {pages} {074} (\bibinfo {year}
  {2016})},\ \Eprint {http://arxiv.org/abs/1512.02647} {arXiv:1512.02647
  [hep-ph]} \BibitemShut {NoStop}%
\bibitem [{\citenamefont {Pierce}\ \emph {et~al.}(2018)\citenamefont {Pierce},
  \citenamefont {Shakya}, \citenamefont {Tsai},\ and\ \citenamefont
  {Zhao}}]{Pierce:2017taw}%
  \BibitemOpen
  \bibfield  {author} {\bibinfo {author} {\bibfnamefont {A.}~\bibnamefont
  {Pierce}}, \bibinfo {author} {\bibfnamefont {B.}~\bibnamefont {Shakya}},
  \bibinfo {author} {\bibfnamefont {Y.}~\bibnamefont {Tsai}}, \ and\ \bibinfo
  {author} {\bibfnamefont {Y.}~\bibnamefont {Zhao}},\ }\href {\doibase
  10.1103/PhysRevD.97.095033} {\bibfield  {journal} {\bibinfo  {journal} {Phys.
  Rev. D}\ }\textbf {\bibinfo {volume} {97}},\ \bibinfo {pages} {095033}
  (\bibinfo {year} {2018})},\ \Eprint {http://arxiv.org/abs/1708.05389}
  {arXiv:1708.05389 [hep-ph]} \BibitemShut {NoStop}%
\bibitem [{\citenamefont {Burdman}\ and\ \citenamefont
  {Lichtenstein}(2018)}]{Lichtenstein:2018kno}%
  \BibitemOpen
  \bibfield  {author} {\bibinfo {author} {\bibfnamefont {G.}~\bibnamefont
  {Burdman}}\ and\ \bibinfo {author} {\bibfnamefont {G.}~\bibnamefont
  {Lichtenstein}},\ }\href {\doibase 10.1007/JHEP08(2018)146} {\bibfield
  {journal} {\bibinfo  {journal} {JHEP}\ }\textbf {\bibinfo {volume} {08}},\
  \bibinfo {pages} {146} (\bibinfo {year} {2018})},\ \Eprint
  {http://arxiv.org/abs/1807.03801} {arXiv:1807.03801 [hep-ph]} \BibitemShut
  {NoStop}%
\bibitem [{\citenamefont {Kilic}\ \emph {et~al.}(2019)\citenamefont {Kilic},
  \citenamefont {Najjari},\ and\ \citenamefont {Verhaaren}}]{Kilic:2018sew}%
  \BibitemOpen
  \bibfield  {author} {\bibinfo {author} {\bibfnamefont {C.}~\bibnamefont
  {Kilic}}, \bibinfo {author} {\bibfnamefont {S.}~\bibnamefont {Najjari}}, \
  and\ \bibinfo {author} {\bibfnamefont {C.~B.}\ \bibnamefont {Verhaaren}},\
  }\href {\doibase 10.1103/PhysRevD.99.075029} {\bibfield  {journal} {\bibinfo
  {journal} {Phys. Rev. D}\ }\textbf {\bibinfo {volume} {99}},\ \bibinfo
  {pages} {075029} (\bibinfo {year} {2019})},\ \Eprint
  {http://arxiv.org/abs/1812.08173} {arXiv:1812.08173 [hep-ph]} \BibitemShut
  {NoStop}%
\bibitem [{\citenamefont {Alipour-Fard}\ \emph {et~al.}(2020)\citenamefont
  {Alipour-Fard}, \citenamefont {Craig}, \citenamefont {Gori}, \citenamefont
  {Koren},\ and\ \citenamefont {Redigolo}}]{Alipour-fard:2018mre}%
  \BibitemOpen
  \bibfield  {author} {\bibinfo {author} {\bibfnamefont {S.}~\bibnamefont
  {Alipour-Fard}}, \bibinfo {author} {\bibfnamefont {N.}~\bibnamefont {Craig}},
  \bibinfo {author} {\bibfnamefont {S.}~\bibnamefont {Gori}}, \bibinfo {author}
  {\bibfnamefont {S.}~\bibnamefont {Koren}}, \ and\ \bibinfo {author}
  {\bibfnamefont {D.}~\bibnamefont {Redigolo}},\ }\href {\doibase
  10.1007/JHEP07(2020)029} {\bibfield  {journal} {\bibinfo  {journal} {JHEP}\
  }\textbf {\bibinfo {volume} {07}},\ \bibinfo {pages} {029} (\bibinfo {year}
  {2020})},\ \Eprint {http://arxiv.org/abs/1812.09315} {arXiv:1812.09315
  [hep-ph]} \BibitemShut {NoStop}%
\bibitem [{\citenamefont {Li}\ and\ \citenamefont {Tsai}(2019)}]{Li:2019ulz}%
  \BibitemOpen
  \bibfield  {author} {\bibinfo {author} {\bibfnamefont {L.}~\bibnamefont
  {Li}}\ and\ \bibinfo {author} {\bibfnamefont {Y.}~\bibnamefont {Tsai}},\
  }\href {\doibase 10.1007/JHEP05(2019)072} {\bibfield  {journal} {\bibinfo
  {journal} {JHEP}\ }\textbf {\bibinfo {volume} {05}},\ \bibinfo {pages} {072}
  (\bibinfo {year} {2019})},\ \Eprint {http://arxiv.org/abs/1901.09936}
  {arXiv:1901.09936 [hep-ph]} \BibitemShut {NoStop}%
\bibitem [{\citenamefont {Buttazzo}\ \emph {et~al.}(2015)\citenamefont
  {Buttazzo}, \citenamefont {Sala},\ and\ \citenamefont
  {Tesi}}]{Buttazzo:2015bka}%
  \BibitemOpen
  \bibfield  {author} {\bibinfo {author} {\bibfnamefont {D.}~\bibnamefont
  {Buttazzo}}, \bibinfo {author} {\bibfnamefont {F.}~\bibnamefont {Sala}}, \
  and\ \bibinfo {author} {\bibfnamefont {A.}~\bibnamefont {Tesi}},\ }\href
  {\doibase 10.1007/JHEP11(2015)158} {\bibfield  {journal} {\bibinfo  {journal}
  {JHEP}\ }\textbf {\bibinfo {volume} {11}},\ \bibinfo {pages} {158} (\bibinfo
  {year} {2015})},\ \Eprint {http://arxiv.org/abs/1505.05488} {arXiv:1505.05488
  [hep-ph]} \BibitemShut {NoStop}%
\bibitem [{\citenamefont {Chacko}\ \emph {et~al.}(2018)\citenamefont {Chacko},
  \citenamefont {Kilic}, \citenamefont {Najjari},\ and\ \citenamefont
  {Verhaaren}}]{Chacko:2017xpd}%
  \BibitemOpen
  \bibfield  {author} {\bibinfo {author} {\bibfnamefont {Z.}~\bibnamefont
  {Chacko}}, \bibinfo {author} {\bibfnamefont {C.}~\bibnamefont {Kilic}},
  \bibinfo {author} {\bibfnamefont {S.}~\bibnamefont {Najjari}}, \ and\
  \bibinfo {author} {\bibfnamefont {C.~B.}\ \bibnamefont {Verhaaren}},\ }\href
  {\doibase 10.1103/PhysRevD.97.055031} {\bibfield  {journal} {\bibinfo
  {journal} {Phys. Rev. D}\ }\textbf {\bibinfo {volume} {97}},\ \bibinfo
  {pages} {055031} (\bibinfo {year} {2018})},\ \Eprint
  {http://arxiv.org/abs/1711.05300} {arXiv:1711.05300 [hep-ph]} \BibitemShut
  {NoStop}%
\bibitem [{\citenamefont {Bishara}\ and\ \citenamefont
  {Verhaaren}(2019)}]{Bishara:2018sgl}%
  \BibitemOpen
  \bibfield  {author} {\bibinfo {author} {\bibfnamefont {F.}~\bibnamefont
  {Bishara}}\ and\ \bibinfo {author} {\bibfnamefont {C.~B.}\ \bibnamefont
  {Verhaaren}},\ }\href {\doibase 10.1007/JHEP05(2019)016} {\bibfield
  {journal} {\bibinfo  {journal} {JHEP}\ }\textbf {\bibinfo {volume} {05}},\
  \bibinfo {pages} {016} (\bibinfo {year} {2019})},\ \Eprint
  {http://arxiv.org/abs/1811.05977} {arXiv:1811.05977 [hep-ph]} \BibitemShut
  {NoStop}%
\bibitem [{\citenamefont {Cheng}\ \emph {et~al.}(2017)\citenamefont {Cheng},
  \citenamefont {Salvioni},\ and\ \citenamefont {Tsai}}]{Cheng:2016uqk}%
  \BibitemOpen
  \bibfield  {author} {\bibinfo {author} {\bibfnamefont {H.-C.}\ \bibnamefont
  {Cheng}}, \bibinfo {author} {\bibfnamefont {E.}~\bibnamefont {Salvioni}}, \
  and\ \bibinfo {author} {\bibfnamefont {Y.}~\bibnamefont {Tsai}},\ }\href
  {\doibase 10.1103/PhysRevD.95.115035} {\bibfield  {journal} {\bibinfo
  {journal} {Phys. Rev. D}\ }\textbf {\bibinfo {volume} {95}},\ \bibinfo
  {pages} {115035} (\bibinfo {year} {2017})},\ \Eprint
  {http://arxiv.org/abs/1612.03176} {arXiv:1612.03176 [hep-ph]} \BibitemShut
  {NoStop}%
\bibitem [{\citenamefont {ATLAS}(2021)}]{ATLAS:2021flf}%
  \BibitemOpen
  \bibfield  {author} {\bibinfo {author} {\bibnamefont {ATLAS}},\ }\href@noop
  {} {\  (\bibinfo {year} {2021})},\ \Eprint
  {http://arxiv.org/abs/ATLAS-CONF-2021-005} {ATLAS-CONF-2021-005} \BibitemShut
  {NoStop}%
\bibitem [{\citenamefont {Sirunyan}\ \emph {et~al.}(2020)\citenamefont
  {Sirunyan} \emph {et~al.}}]{Sirunyan:2020cao}%
  \BibitemOpen
  \bibfield  {author} {\bibinfo {author} {\bibfnamefont {A.~M.}\ \bibnamefont
  {Sirunyan}} \emph {et~al.} (\bibinfo {collaboration} {CMS}),\ }\href@noop {}
  {\  (\bibinfo {year} {2020})},\ \Eprint {http://arxiv.org/abs/2012.01581}
  {arXiv:2012.01581 [hep-ex]} \BibitemShut {NoStop}%
\bibitem [{\citenamefont {CMS}(2021{\natexlab{a}})}]{CMS:2021uxj}%
  \BibitemOpen
  \bibfield  {author} {\bibinfo {author} {\bibnamefont {CMS}},\ }\href@noop {}
  {\  (\bibinfo {year} {2021}{\natexlab{a}})},\ \Eprint
  {http://arxiv.org/abs/CMS-PAS-EXO-20-003} {CMS-PAS-EXO-20-003} \BibitemShut
  {NoStop}%
\bibitem [{\citenamefont {Aaboud}\ \emph
  {et~al.}(2019{\natexlab{a}})\citenamefont {Aaboud} \emph
  {et~al.}}]{Aaboud:2018aqj}%
  \BibitemOpen
  \bibfield  {author} {\bibinfo {author} {\bibfnamefont {M.}~\bibnamefont
  {Aaboud}} \emph {et~al.} (\bibinfo {collaboration} {ATLAS}),\ }\href
  {\doibase 10.1103/PhysRevD.99.052005} {\bibfield  {journal} {\bibinfo
  {journal} {Phys. Rev. D}\ }\textbf {\bibinfo {volume} {99}},\ \bibinfo
  {pages} {052005} (\bibinfo {year} {2019}{\natexlab{a}})},\ \Eprint
  {http://arxiv.org/abs/1811.07370} {arXiv:1811.07370 [hep-ex]} \BibitemShut
  {NoStop}%
\bibitem [{\citenamefont {Aaboud}\ \emph
  {et~al.}(2019{\natexlab{b}})\citenamefont {Aaboud} \emph
  {et~al.}}]{Aaboud:2019opc}%
  \BibitemOpen
  \bibfield  {author} {\bibinfo {author} {\bibfnamefont {M.}~\bibnamefont
  {Aaboud}} \emph {et~al.} (\bibinfo {collaboration} {ATLAS}),\ }\href
  {\doibase 10.1140/epjc/s10052-019-6962-6} {\bibfield  {journal} {\bibinfo
  {journal} {Eur. Phys. J. C}\ }\textbf {\bibinfo {volume} {79}},\ \bibinfo
  {pages} {481} (\bibinfo {year} {2019}{\natexlab{b}})},\ \Eprint
  {http://arxiv.org/abs/1902.03094} {arXiv:1902.03094 [hep-ex]} \BibitemShut
  {NoStop}%
\bibitem [{\citenamefont {CMS}(2021{\natexlab{b}})}]{CMS:2021zdu}%
  \BibitemOpen
  \bibfield  {author} {\bibinfo {author} {\bibnamefont {CMS}},\ }\href@noop {}
  {\  (\bibinfo {year} {2021}{\natexlab{b}})},\ \Eprint
  {http://arxiv.org/abs/CMS-PAS-EXO-20-015} {CMS-PAS-EXO-20-015} \BibitemShut
  {NoStop}%
\bibitem [{\citenamefont {Vilenkin}(1981)}]{Vilenkin:1981zs}%
  \BibitemOpen
  \bibfield  {author} {\bibinfo {author} {\bibfnamefont {A.}~\bibnamefont
  {Vilenkin}},\ }\href {\doibase 10.1103/PhysRevD.23.852} {\bibfield  {journal}
  {\bibinfo  {journal} {Phys. Rev. D}\ }\textbf {\bibinfo {volume} {23}},\
  \bibinfo {pages} {852} (\bibinfo {year} {1981})}\BibitemShut {NoStop}%
\bibitem [{\citenamefont {Juknevich}\ \emph {et~al.}(2009)\citenamefont
  {Juknevich}, \citenamefont {Melnikov},\ and\ \citenamefont
  {Strassler}}]{Juknevich:2009ji}%
  \BibitemOpen
  \bibfield  {author} {\bibinfo {author} {\bibfnamefont {J.~E.}\ \bibnamefont
  {Juknevich}}, \bibinfo {author} {\bibfnamefont {D.}~\bibnamefont {Melnikov}},
  \ and\ \bibinfo {author} {\bibfnamefont {M.~J.}\ \bibnamefont {Strassler}},\
  }\href {\doibase 10.1088/1126-6708/2009/07/055} {\bibfield  {journal}
  {\bibinfo  {journal} {JHEP}\ }\textbf {\bibinfo {volume} {07}},\ \bibinfo
  {pages} {055} (\bibinfo {year} {2009})},\ \Eprint
  {http://arxiv.org/abs/0903.0883} {arXiv:0903.0883 [hep-ph]} \BibitemShut
  {NoStop}%
\bibitem [{\citenamefont {Juknevich}(2010{\natexlab{a}})}]{Juknevich:2009gg}%
  \BibitemOpen
  \bibfield  {author} {\bibinfo {author} {\bibfnamefont {J.~E.}\ \bibnamefont
  {Juknevich}},\ }\href {\doibase 10.1007/JHEP08(2010)121} {\bibfield
  {journal} {\bibinfo  {journal} {JHEP}\ }\textbf {\bibinfo {volume} {08}},\
  \bibinfo {pages} {121} (\bibinfo {year} {2010}{\natexlab{a}})},\ \Eprint
  {http://arxiv.org/abs/0911.5616} {arXiv:0911.5616 [hep-ph]} \BibitemShut
  {NoStop}%
\bibitem [{Note3()}]{Note3}%
  \BibitemOpen
  \bibinfo {note} {This assumes that a sizeable fraction of produced glueballs
  are the shortest-lived lightest $0^{++}$ state, which is backed up by
  finite-temperature QCD estimates~\cite {Juknevich:2010rhj}}\BibitemShut
  {NoStop}%
\bibitem [{\citenamefont {Curtin}\ \emph {et~al.}(2018)\citenamefont {Curtin}
  \emph {et~al.}}]{Curtin:2018mvb}%
  \BibitemOpen
  \bibfield  {author} {\bibinfo {author} {\bibfnamefont {D.}~\bibnamefont
  {Curtin}} \emph {et~al.},\ }\href@noop {} {\  (\bibinfo {year} {2018})},\
  \Eprint {http://arxiv.org/abs/1806.07396} {arXiv:1806.07396 [hep-ph]}
  \BibitemShut {NoStop}%
%%CITATION = ARXIV:1806.07396;%%
\bibitem [{\citenamefont {Alpigiani}\ \emph {et~al.}(2020)\citenamefont
  {Alpigiani} \emph {et~al.}}]{Alpigiani:2020tva}%
  \BibitemOpen
  \bibfield  {author} {\bibinfo {author} {\bibfnamefont {C.}~\bibnamefont
  {Alpigiani}} \emph {et~al.} (\bibinfo {collaboration} {MATHUSLA}),\
  }\href@noop {} {\  (\bibinfo {year} {2020})},\ \Eprint
  {http://arxiv.org/abs/2009.01693} {arXiv:2009.01693 [physics.ins-det]}
  \BibitemShut {NoStop}%
\bibitem [{\citenamefont {Ito}\ \emph {et~al.}(2017)\citenamefont {Ito},
  \citenamefont {Jinnouchi}, \citenamefont {Moroi}, \citenamefont {Nagata},\
  and\ \citenamefont {Otono}}]{Ito:2017dpm}%
  \BibitemOpen
  \bibfield  {author} {\bibinfo {author} {\bibfnamefont {H.}~\bibnamefont
  {Ito}}, \bibinfo {author} {\bibfnamefont {O.}~\bibnamefont {Jinnouchi}},
  \bibinfo {author} {\bibfnamefont {T.}~\bibnamefont {Moroi}}, \bibinfo
  {author} {\bibfnamefont {N.}~\bibnamefont {Nagata}}, \ and\ \bibinfo {author}
  {\bibfnamefont {H.}~\bibnamefont {Otono}},\ }\href {\doibase
  10.1016/j.physletb.2017.06.003} {\bibfield  {journal} {\bibinfo  {journal}
  {Phys. Lett. B}\ }\textbf {\bibinfo {volume} {771}},\ \bibinfo {pages} {568}
  (\bibinfo {year} {2017})},\ \Eprint {http://arxiv.org/abs/1702.08613}
  {arXiv:1702.08613 [hep-ph]} \BibitemShut {NoStop}%
\bibitem [{\citenamefont {Ito}\ \emph {et~al.}(2018)\citenamefont {Ito},
  \citenamefont {Jinnouchi}, \citenamefont {Moroi}, \citenamefont {Nagata},\
  and\ \citenamefont {Otono}}]{Ito:2018asa}%
  \BibitemOpen
  \bibfield  {author} {\bibinfo {author} {\bibfnamefont {H.}~\bibnamefont
  {Ito}}, \bibinfo {author} {\bibfnamefont {O.}~\bibnamefont {Jinnouchi}},
  \bibinfo {author} {\bibfnamefont {T.}~\bibnamefont {Moroi}}, \bibinfo
  {author} {\bibfnamefont {N.}~\bibnamefont {Nagata}}, \ and\ \bibinfo {author}
  {\bibfnamefont {H.}~\bibnamefont {Otono}},\ }\href {\doibase
  10.1007/JHEP06(2018)112} {\bibfield  {journal} {\bibinfo  {journal} {JHEP}\
  }\textbf {\bibinfo {volume} {06}},\ \bibinfo {pages} {112} (\bibinfo {year}
  {2018})},\ \Eprint {http://arxiv.org/abs/1803.00234} {arXiv:1803.00234
  [hep-ph]} \BibitemShut {NoStop}%
\bibitem [{\citenamefont {Liu}\ \emph {et~al.}(2017)\citenamefont {Liu},
  \citenamefont {Wang},\ and\ \citenamefont {Zhang}}]{Liu:2016zki}%
  \BibitemOpen
  \bibfield  {author} {\bibinfo {author} {\bibfnamefont {Z.}~\bibnamefont
  {Liu}}, \bibinfo {author} {\bibfnamefont {L.-T.}\ \bibnamefont {Wang}}, \
  and\ \bibinfo {author} {\bibfnamefont {H.}~\bibnamefont {Zhang}},\ }\href
  {\doibase 10.1088/1674-1137/41/6/063102} {\bibfield  {journal} {\bibinfo
  {journal} {Chin. Phys. C}\ }\textbf {\bibinfo {volume} {41}},\ \bibinfo
  {pages} {063102} (\bibinfo {year} {2017})},\ \Eprint
  {http://arxiv.org/abs/1612.09284} {arXiv:1612.09284 [hep-ph]} \BibitemShut
  {NoStop}%
\bibitem [{\citenamefont {Alipour-Fard}\ \emph {et~al.}(2019)\citenamefont
  {Alipour-Fard}, \citenamefont {Craig}, \citenamefont {Jiang},\ and\
  \citenamefont {Koren}}]{Alipour-Fard:2018lsf}%
  \BibitemOpen
  \bibfield  {author} {\bibinfo {author} {\bibfnamefont {S.}~\bibnamefont
  {Alipour-Fard}}, \bibinfo {author} {\bibfnamefont {N.}~\bibnamefont {Craig}},
  \bibinfo {author} {\bibfnamefont {M.}~\bibnamefont {Jiang}}, \ and\ \bibinfo
  {author} {\bibfnamefont {S.}~\bibnamefont {Koren}},\ }\href {\doibase
  10.1088/1674-1137/43/5/053101} {\bibfield  {journal} {\bibinfo  {journal}
  {Chin. Phys. C}\ }\textbf {\bibinfo {volume} {43}},\ \bibinfo {pages}
  {053101} (\bibinfo {year} {2019})},\ \Eprint
  {http://arxiv.org/abs/1812.05588} {arXiv:1812.05588 [hep-ph]} \BibitemShut
  {NoStop}%
\bibitem [{\citenamefont {Cheung}\ and\ \citenamefont
  {Wang}(2020)}]{Cheung:2019qdr}%
  \BibitemOpen
  \bibfield  {author} {\bibinfo {author} {\bibfnamefont {K.}~\bibnamefont
  {Cheung}}\ and\ \bibinfo {author} {\bibfnamefont {Z.~S.}\ \bibnamefont
  {Wang}},\ }\href {\doibase 10.1103/PhysRevD.101.035003} {\bibfield  {journal}
  {\bibinfo  {journal} {Phys. Rev. D}\ }\textbf {\bibinfo {volume} {101}},\
  \bibinfo {pages} {035003} (\bibinfo {year} {2020})},\ \Eprint
  {http://arxiv.org/abs/1911.08721} {arXiv:1911.08721 [hep-ph]} \BibitemShut
  {NoStop}%
\bibitem [{\citenamefont {Juknevich}(2010{\natexlab{b}})}]{Juknevich:2010rhj}%
  \BibitemOpen
  \bibfield  {author} {\bibinfo {author} {\bibfnamefont {J.~E.}\ \bibnamefont
  {Juknevich}},\ }\emph {\bibinfo {title} {{Phenomenology of pure-gauge hidden
  valleys at Hadron colliders}}},\ \href {\doibase 10.7282/T34F1QHM} {Ph.D.
  thesis},\ \bibinfo  {school} {Rutgers U., Piscataway} (\bibinfo {year}
  {2010}{\natexlab{b}})\BibitemShut {NoStop}%
\end{thebibliography}%

\end{document}